\documentclass[article]{aa}  

\usepackage{graphicx}
\usepackage{txfonts}
\begin{document}

   \title{A Virgo Environmental Survey Tracing Ionised Gas Emission. VESTIGE VIII. Bridging the cluster-ICM-galaxy evolution at small scales }

   \author{A.~Longobardi\thanks{AL acknowledges support from the French Centre National d’Etudes Spatiales (CNES)}
          \inst{1}
          \and
          A.~Boselli\inst{1}
          \and
          M.~Fossati\inst{2}
          \and
          J.~A.~Villa-V\'{e}lez
          \inst{1}
          \and
          S.~Bianchi\inst{3}
          \and
          V.~Casasola\inst{4}
          \and
          E.~Sarpa\inst{5,6}
          \and
          F.~Combes\inst{7}
          \and
          G.~Hensler\inst{8}
          \and
          D.~Burgarella\inst{1}
          \and 
          C.~Schimd\inst{1}
          \and
          A.~Nanni\inst{1}
          \and
          P.~C\^{o}t\'{e}\inst{9}
          \and
          V.~Buat\inst{1}
          \and
          P.~Amram\inst{1}
          \and
          L.~Ferrarese\inst{9}
          \and
          J.~Braine\inst{10}
          \and
          G.~Trinchieri\inst{11}
          \and
          S.~Boissier\inst{1}
          \and
          M.~Boquien\inst{12}
          \and
          P.~Andreani\inst{13}
          \and
          S.~Gwyn\inst{8}
          \and
          J.~C.~Cuillandre\inst{14}
          }

   \institute{Aix Marseille Univ, CNRS, CNES, LAM, Marseille, France\\
              \email{alessia.longobardi@lam.fr}
         \and
             Dipartimento di Fisica G. Occhialini, Universit\'{a} degli Studi di Milano Bicocca, Piazza della Scienza 3, I-20126 Milano, Italy
        \and
             INAF – Osservatorio Astrofisico di Arcetri, Largo E. Fermi 5, 50125, Firenze, Italy
        \and
            INAF – Istituto di Radioastronomia, Via P. Gobetti 101, 40129, Bologna, Italy
        \and
        Dipartimento di Matematica e Fisica, Universit\'{a} degli studi Roma Tre, Via della Vasca Navale, 84, 00146 Roma, Italy 
        \and 
        INFN - Sezione di Roma Tre, via della Vasca Navale 84, I-00146 Roma, Italy
        \and 
        Observatoire de Paris, Coll\'{e}ge de France, PSL University, Sorbonne University, CNRS, LERMA, Paris
        \and
        Department for Astrophysics, University of Vienna, T\"{u}rkenschanzstrasse 17, A-1180 Vienna, Austria
        \and
        National Research Council of Canada, Herzberg Astronomy \& Astrophysics Research Centre, 5071 W. Saanich Rd, Victoria, BC, V9E
        \and
        Laboratoire d'Astrophysique de Bordeaux, Univ. Bordeaux, CNRS, B18N, Allée Geoffroy Saint-Hilaire, 33615 Pessac, France
        \and
        INAF - Osservatorio Astronomico di Brera, Via Brera 28, 20159 Milano, Italy
        \and
            European Southern Observatory, Karl-Schwarzschild-Strasse 2, 85748 Garching, Germany
        \and
            7 AIM, CEA, CNRS, Universit\'{e} Paris-Saclay, Universit\'{e}  Paris Diderot, Sorbonne Paris Cit\'{e}, Observatoire de Paris,
            PSL University, 91191 Gif-sur-Yvette Cedex, France
             } 

   \date{}

 
  \abstract
   {}
   { We measure FIR emission from tails of stripped dust following the ionised and atomic gas components in galaxies undergoing ram pressure stripping. We study the dust-to-gas relative distribution and mass  ratio in the stripped interstellar medium and relate them to those of the intra-cluster medium, thus linking the cluster-ICM-galaxy evolution at small-scale.
   The galaxy sample consists of three Scd Virgo galaxies with stellar masses in the range $10^9\lesssim \mathrm{M_{*}} \lesssim 10^{10}\, \mathrm{M_{\odot}}$, and within 1 Mpc from the cluster centre, namely NGC 4330, NGC 4522, and NGC 4654.}
   {Through the analysis of VESTIGE H$\alpha$, $Herschel$ SPIRE far-infrared, and VIVA HI data, we trace the spatial distribution of the tails and infer the dust and gas masses from the measured far-infrared 250 $\mu$m and HI flux densities. Dust-to-gas mass ratios in the tails are analysed as a function of the galaxy mass, metallicity, and dust temperature.}
   {Along the stripped component, the dust distribution closely follows the HI and H$\alpha$ emitting gas, all extending beyond the optical disc  (defined by the $B$-band $25^{th}$ magnitude isophote). 
   In these regions the dust-to-gas mass ratios are 
   $2.0\pm0.6\times10^{-3}$, $0.7\pm0.1\times10^{-3}$, and $0.4\pm0.03\times10^{-3}$, for NGC 4330, NGC 4522, and NGC 4654, respectively. Thus, dust is widespread in the stripped material with a lower dust-to-gas mass ratio (up to a factor of 15) than the one measured in the main body of nearby galaxies. We also find a negative trend in the dust-to-gas mass ratio as a function of the metallicity that can be explained in terms of a dust component more centrally concentrated in more metal-rich systems. Together with the finding that the stripped dust is cold, $T_{d} \lesssim 25\, K$, our results can be interpreted as a consequence of an outside-in stripping of the galaxy interstellar medium.}
   { Gas and dust in galaxies are perturbed in a similar fashion by the cluster environment, although their relative contribution differs from the one measured in the main body of the galaxies. When this value is considered, ram pressure stripping is consistent with being one of the key mechanisms in the building up of the Virgo intra-cluster component, injecting dust grains into the ICM, thus contributing to its metal enrichment.}

   \maketitle
%
\section{Introduction}
In a $\Lambda$CDM Universe, where  the hierarchical evolution is the driving mechanism in determining the current epoch characteristics of galaxies, it is expected that an abundance of low surface brightness, intra-cluster tidal debris from disrupted systems, and an ubiquity of diffuse structures, permeate the intra-cluster medium (ICM) of galaxy clusters \citep[e.g.][]{napolitano03,murante04,sommer05}.
In fact, as a consequence of environmental processing, when a galaxy enters a high-density region it can be subjected to gravitational interactions with other galaxies or with the potential well of the over-dense region, or it can feel the pressure exerted by the hot and dense ICM \citep{boselligavazzi06}. Baryons are stripped from the main body of the galaxies resulting in the production of tails of stripped material. This history is often hidden, however, being at surface brightness levels much fainter than the sky. It is only thanks to advances in sensitivity, angular, and spatial resolution of modern multi-frequency instrumentation that astronomers have been able to collect growing evidence of objects undergoing stripping of the different phases of the inter stellar medium (ISM). Long tails of atomic gas have been detected and interpreted as the result of the stripping of the (more extended) galaxy HI disk \citep[e.g.,][]{chung09}. In some cases, this gas also appears as ionised (hotter) and detected through its H$\alpha$ emission
\citep[e.g.,][]{boselli16,poggianti17,fossati18,bellhouse19},
or it may get heated to the cluster X-ray temperature \citep{sun06}. 
Finally, cold molecular gas, usually detected through its CO emission, has been found to follow the stripped HI component  \citep[e.g.][]{vollmer08,verdugo15,jachym17,moretti18,cramer19, zabel19,moretti20}. \\
\indent
Since the dust is also perturbed by environmental processing \citep[e.g.][]{cortese10a,cortese10b,kenney15,abramson16}
and in galaxies it is associated with the gaseous component of the ISM, it is generally expected that part of the dust is removed together with the gas during stripping. Several studies in the literature have identified dust in absorption in stripped tails through the analysis of the Balmer decrement. However, since it is associated to extra-planar HII star-forming regions, this is likely formed \textit{in situ} \citep[e.g][]{fossati16,poggianti17,gullieuszik17, poggianti19, bianconi20}.
Thus, to which extent the dust follows the same fate as the stripped hydrogen is still an open question. Is the dust-to-gas ratios measured in the main body of the galaxies the same as in the tails of systems undergoing environmental processing? How do the properties between the different phases of the ISM change during the different stages of evolution?
Additionally, if the stripped tails lie outside the galaxies' optical disks, they are likely going to be removed 
from the cluster spirals and to build up the cluster intra-cluster component (ICC).\\ 
\indent
This work aims at bridging the cluster-ICM-galaxy evolution at small scales analysing the relative fraction in mass of the different baryonic components in the stripped ISM of galaxies subject to environmental processes. Our study targets galaxies in the Virgo cluster, the nearest large concentration of mass, a dynamically young cluster \citep[e.g.][]{binggeli87,boehringer94} for which an exceptional collection of multi-frequency data at good/optimal resolution and sensitivity is available. In this context, the Virgo Environmental Survey Tracing Ionised Gas Emission \citep[VESTIGE;][]{boselli18} is a new, blind H$\alpha$ survey providing us with the largest and deepest information on the ionised gas emission in Virgo, revealing unknown tails of stripped gas in several cluster members.
Additionally, the Virgo cluster has recently been shown to contain a diffuse IC dust (ICD) component that is widespread in the cluster ICM with a dust-to-gas mass ratio $M_{\mathrm{d}}/M_{\mathrm{g}} = 3.0 \pm 0.3 \times 10^{-4}$ likely transported into the IC space by stripping \citep{longobardi20}. \\
\noindent

Throughout the paper, we consider the cluster centred on M87, with a virial radius $r_{\mathrm{vir}}=$  1.55 Mpc \citep{mclaughlin99}.
By assuming a flat $\Lambda$CDM Universe with $\Omega_{\mathrm{M}} = 0.3$, $\Omega_{\Lambda} = 0.7$, and $H_{0} = 70\, \mathrm{km\, s^{-1}\, Mpc^{-1}}$ and a distance for Virgo of 16.5 Mpc \citep{gavazzi99,mei07,blakeslee09},  the adopted physical scale is 80 pc arcsec$^{-1}$. The optical disc extention refers to the $B$-band 25$^{th}$ magnitude isophote.



\section{Photometric data}
\label{data}
The data comes from a compilation of multi-wavelength campaigns surveying the Virgo cluster in the ultra-violet (UV), optical, radio, and far infra-red (FIR). In what follows, we give a brief description of these surveys and refer the reader to the references therein for additional details.
\vspace{-0.5cm}
\paragraph{{\bf The VESTIGE survey}}
VESTIGE is a blind survey of the Virgo cluster carried out with MegaCam on the Canada-France-Hawaii Telescope (CFHT) with the H$\alpha$ narrow band filter \footnote{Given the characteristics of the CFHT H$\alpha $ filter (central wavelength $\lambda_c=6591$ \AA and band width 106 \AA) the VESTIGE data include [NII] line contribution.} and the broad-band $r'$ filter \citep{boselli18}. It is designed to cover a total area of 104 deg$^ 
2$, and reach the two main subclusters (Virgo A centered on M87, and Virgo B centered on M49) out to their virial radii. Currently the survey covers 40~\% of the designed area at full depth with the current observations taken in excellent weather conditions (median seeing $\sim$ 0.62\arcsec and 0.65\arcsec in the narrow-band and $r'$-band filter, respectively). VESTIGE data have been reduced using the Elixir-LSB package \citep{ferrarese12} that is optimised for the 
removal of the instrumental background and scattered light from the science frames. This provides high signal-to-noise ratio of the extended low surface brightness features, making VESTIGE a deep photometric survey that for extended sources reaches a depth of $\Sigma(H\alpha) \sim 2 \times 10^{-18} \mathrm{erg\, sec^{-1} cm^{-2} arcsec^{-2}}$ at 3\arcsec resolution. The photometric zero points were tied to Pan-STARRS photometry for both filters with a final photometric uncertainty of $\sim 2-3~\%$ \citep[see][]{boselli18}. All final images have the same astrometric reference frame, tied to the positions of stars in the Sloan Digital Sky Survey (SDSS), with a spatial scale of 0.186 \arcsec px$^{-1}$ \citep{gwyn08}. Finally, H$\alpha$ images with the only nebular line contribution are obtained via subtraction of the stellar continuum. The latter is obtained scaling the $r'$-band image by a ($g'-r'$) colour factor that accounts for the difference in central wavelength of the narrow and broad band filters \citep[see][]{boselli18,fossati18,boselli19,boselli20}. The optical $g'$-band information is taken from the Next Generation Virgo Cluster Survey \citep{ferrarese12}, that we describe below. 
\vspace{-0.5cm}
\paragraph{{\bf The NGVS survey}}
The broad band optical information is taken from the Next Generation Virgo Cluster Survey \citep[NGVS;][]{ferrarese12}, a deep CFHT program in the $u^*$ $g'$ $i'$ $z'$ bands that covers a total area of 104 deg$^{2}$ in Virgo. The data were reduced with the Elixir-LSB pipeline and the photometric zero points were tied to SDSS photometry, as was done for the VESTIGE data. The typical FWHM (full-width-half-maximum) is $\sim0.55\arcsec$ in the $i'$ band and $\sim0.8\arcsec$ in the other bands. In the $g'$ band, the NGVS reaches a depth for extend sources of $g'= 27.7\, \mathrm{ mag\, arcesc^{2}}$.
\vspace{-0.5cm}
\paragraph{{\bf The HeViCS survey}}
Far-infrared data come from the $Herschel$ Virgo Cluster Survey \citep[HeViCS;][]{davies10}, a programme that covers $\sim$60 deg$^2$ of the Virgo cluster using the PACS \citep{poglitsch10} instrument at 100 and 160 $\mu$m, and the SPIRE \citep{griffin10} instrument at 250, 350, and 500 $\mu$m. Data were integrated into the $Herschel$ Reference Survey \citep[HRS;][]{boselli10}, and their reduction was carried out as described in \citet{ciesla12} and \citet{cortese14}. The sensitivity and FWHMs of the PACS observations are $\sim 6.8$ and $\sim 3.1\, \mathrm{MJy\, sr^{-1}}$ and 7\arcsec and 12\arcsec  at 100 and 160 $\mu$m, respectively, while the sensitivity and 
FWHMs of the SPIRE observations are 
$\sim 1.0\,, 0.8\,, \mathrm{and}\, 1.0$ $\mathrm{MJy\, sr^{-1}}$ and 
$\sim$18\arcsec,$\sim$25\arcsec, and $\sim$36\arcsec at 250, 350, and 500 $\mu$m, respectively. Among these the FIR 250 $\mu$m observations are the only one that allow for a statistically significant measurement of the fluxes in the tail regions, due to a compromise between spatial resolution and depth (see Sect.\ref{photometry}).  
Because of this reason the main photometric analysis in Sect. \ref{stripped_tails} is based only on SPIRE 250 $\mu$m data, for which the adopted beam size value is the pipeline beam solid angle equal to 469.35 arcsec$^2$.
\vspace{-0.5cm}
\paragraph{{\bf The VIVA survey}}
The VIVA Survey (VLA Imaging of Virgo in Atomic gas) is an imaging survey in HI of 53 Virgo late-type galaxies, covering angular distances of $\sim 1-12$ deg ($\sim 0.3-3.5$ Mpc) from the cluster's centre \citep{chung09}. The total HI image, the intensity weighted velocity field, and the velocity dispersion image were produced using the Astronomical Imaging Processing System (AIPS) by taking moments along the frequency axis (the 0th, 1st, and 2nd moment). This resulted in a HI imaging survey with a typical spatial resolution of 15\arcsec and a column density sensitivity of about $3-5 \times 10^{19}\, \mathrm{cm^{-2}}$ (3$\sigma$) per 10 kms~$^{-1}$ channel.  For our sample of objects (see next Sect.) the beam FWHMs are  $26.36\arcsec \times 23.98\arcsec $, $18.88\arcsec \times 15.20\arcsec $, and $16.14\arcsec \times 15.52\arcsec $, for NGC 4330, NGC 4522, and NGC 4654, respectively. 
\vspace{-0.5cm}
\paragraph{{\bf The GUViCS survey}}
The GUViCS survey \citep[GALEX Ultraviolet Virgo Cluster Survey;][] {boselli11} presents GALEX far-UV (FUV) and near-UV (NUV) observations of the Virgo cluster. It combines data from the All-sky Imaging Survey ($\sim~5$\arcsec spatial resolution and single-exposure times of typically 100 s) and the Medium Imaging Survey (MIS; same spatial resolution, but with deeper exposure times of at least 1500 s), processed with the GALEX pipeline \citep{bianchi14}.

\section{The galaxy sample}
\begin{table*}
    \centering
    \caption{Physical properties of the galaxy sample.}
    \vspace{-0.2cm}
    \resizebox{1.\textwidth}{!}{
    \begin{tabular}{cccccccccccccc}
    \hline
    \hline
    ID &RA &DEC  & $D_{25}^{a}$ & $i^{a}$ & $v^{a}$ & $d_{\mathrm{M87}}$& $\log{M_{*}}^{a}$&$\log{M_{\mathrm{dust}}}^{b}$&$\log{M_{\mathrm{HI}}^{c}}$&$L_{H\alpha}$&$S_{250\mu m}$&$S_{HI}$& $\mathrm{def_{HI}^{c}}$\\
     & J2000 & J2000 &  (arcmin) & (deg)& (km s$^{-1}$)& (Mpc) & ($M_{\odot}$)&($M_{\odot}$)&($M_{\odot}$)  & ($10^{40}$ erg s$^{-1}$)& (Jy) & (Jy km s$^{-1})$&\\
     (1)& (2) & (3) & (4) & (5)& (6)& (7) & (8)  & (9)& (10) & (11)&(12)&(13)&(14)\\
     \hline\\[-5pt]
    NGC4330 & 12:23:17.25 & +11:22:04.7  & 5.86 & 90& 1567& 0.6&9.3& 6.9&8.7&$1.1\pm0.02$&$3.0\pm0.5$&39.7$\pm7.4$&$0.80$\\
    NGC4522 & 12:33:39.66 & +09:10:29.5  & 4.04 & 79& 2332& 0.9&9.3 &6.9&8.8 &$ 1.4 \pm 0.03$&$2.9\pm0.3$ &$59.4\pm11.9$&$0.86$\\
    NGC4654 & 12:43:56.58 & +13:07:36.0 & 4.99 & 56& 1035& 0.9&10.2 &7.8& 9.5&$19.2\pm0.01$&$24.8\pm0.9$ &$73.3\pm14.7$&$0.12$\\
       \hline
    \end{tabular}
    }
    \tablefoot{{\bf Column 1}: Galaxy name. {\bf Column 2-3}: J2000 coordinates. {\bf Column 4}: Optical size defined by the B-band 25$^{th}$ magnitude isophote. {\bf Column 5}: Inclination angle. {\bf Column 6}: Velocity. {\bf Column 7}: Projected distance from M87.{ \bf Columns 8-10}: log values of stellar, dust, and HI gas masses. {\bf Column 11}: Total luminosity in H$\alpha$ within $D_{25}$. {\bf Columns 12-13}: Total FIR 250 $\mu m$, and HI flux densities within $D_{25}$.
    {\bf Column 14}: $HI-def$ parameter measured as the logarithmic difference between the expected and observed HI masses. 
    $^{\bf a}$ \citet{cortese12}.
    $^{\bf b}$ \citet{ciesla12}.
    $^{\bf c}$ \citet{chung09}. $M_{HI}$ values have been scaled to take into account the different distances assumed for the Virgo cluster. }
    \label{tab:physical_p}
\end{table*}
To study the interplay between gas and dust during a late stage of galaxy evolution and its connection with the building up of the Virgo ICC, our study samples galaxies with tails of ionised H$\alpha$, neutral HI and FIR emission, extending beyond the galaxy optical disc, namely NGC 4330, NGC 4522, and NGC 4654.
They are all galaxies of Scd morphological type, located within 4 deg ($\sim 1$ Mpc) from the cluster centre, and with intermediate stellar masses in the range $10^9\lesssim M_{*} \lesssim 10^{10}\, \mathrm{M_\odot}$. Table\ref{tab:physical_p} lists some of the physical properties of the galaxy sample.\\
\indent
NGC 4330 shows truncated disks in UV and H$\alpha$ \citep[e.g.][; Vollmer et al. 2020 in press]{fossati18}, FIR \citep{cortese10a}, HI \citep{chung09,abramson11}, and CO \citep{lee17} on the north-east side of the stellar disc, and low surface-brightness, extended tails of ionised and neutral atomic gas on the southern side. It is a clear example of a galaxy undergoing ram pressure stripping that is effectively quenching the star formation activity with an out-in radial gradient \citep{fossati18}.
NGC 4522 is farther away in projected distance from the centre of the cluster (0.9 Mpc), however still experiencing ram pressure stripping as indicated by the HI and CO asymmetric morphology \citep{vollmer06,vollmer08,chung09}. Finally,  NGC 4654, at the same distance of NGC 4522, shows HI and CO gas distributions compressed in the north-west, but very extended HI gas on the opposite side \citep[CO observation do not extend at such distances;][]{chung14}. The stellar and H$\alpha$ morphologies are also asymmetric, showing an enhancement of ionised emission at the north-west, representative of recent star formation in this region \citep{chung07}, as well as tails of stripped stars in south-est. For these characteristics NGC 4654 may be the only case in our sample of galaxies undergoing both ram pressure and tidal stripping, as also suggested by theoretical models of \citet{vollmer03}.\\
\indent
We stress that this sample is not complete. NGC 4330, NGC 4522, and NGC 4654 represent $40\, \%$ of the Virgo galaxies with H$\alpha$, HI and FIR tails, and only $\sim5\, \%$ of galaxies expected to be subject to ram pressure stripping in Virgo \citep{boselligavazzi14}.
There are three main factors that led to this incompleteness: i) the VESTIGE survey has reached full sensitivity only in the central 5 degrees of the cluster, thus it does not allow for a complete comparison with the VIVA sample, the latter extending out to the edge of the cluster; ii) our sample is biased towards bright and massive galaxies: only 10 VIVA targets are classified as Sd/Sm/Im galaxies and 50~\% of these lie outside the VESTIGE complete region. iii) for the remaining fraction of low mass galaxies within the sampled area the current sensitivity and resolution of the FIR observations are likely prohibitive to detect stripped dust tails. However, these objects are expected to be the most affected by environmental processes.

Limited in statistics, our work is to be considered a pilot study for future campaigns.




\section{Stripped tails}

\label{stripped_tails}
\begin{figure*}
    \centering
    \includegraphics[width=1.\textwidth,trim={2.cm 1.5cm 2cm 0cm},clip]{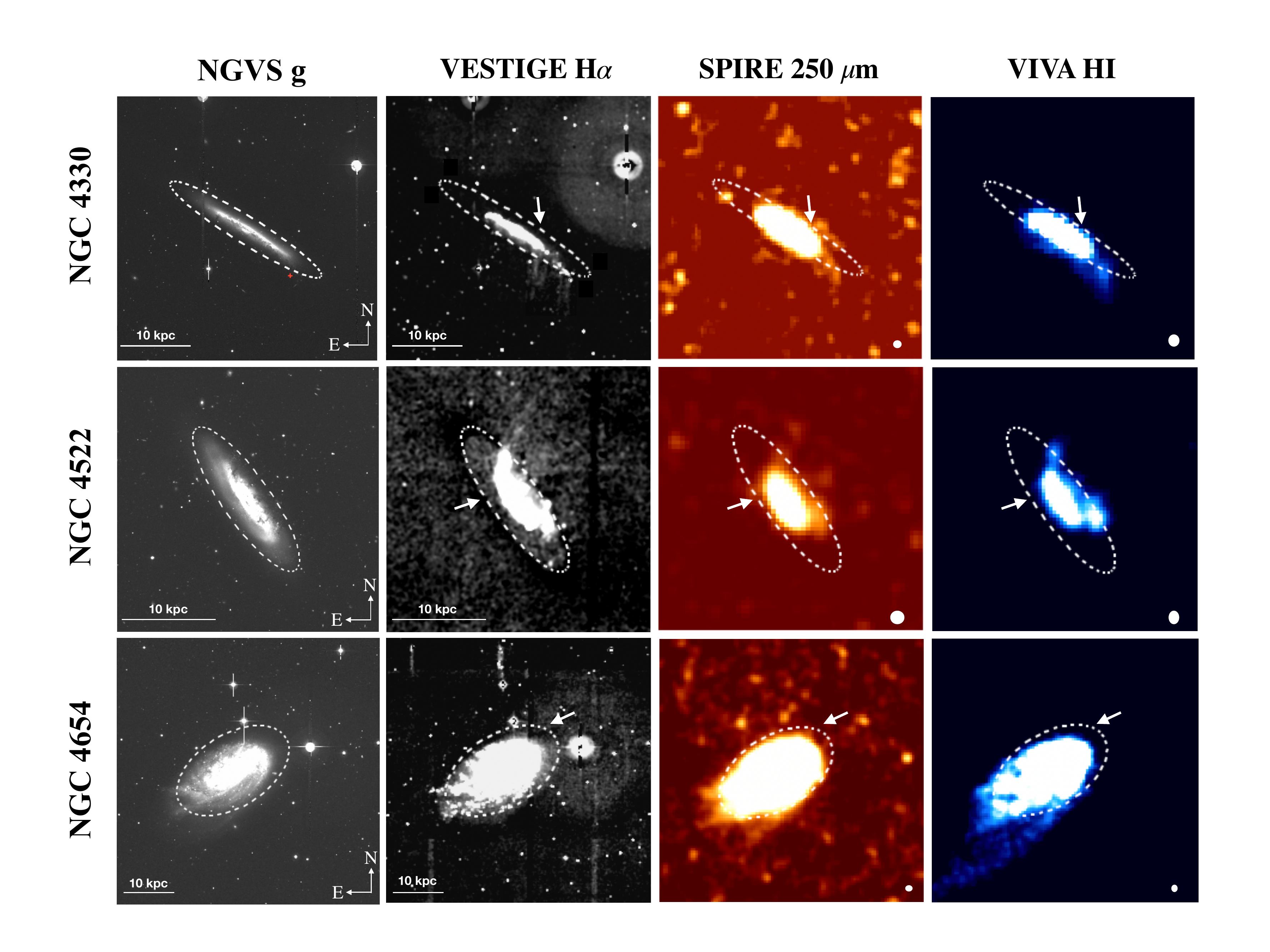}
    \caption{The $g'$-band, H$\alpha$, FIR 250 $\mu$m and HI maps for the galaxies in our sample. The H$\alpha$ and FIR 250$\mu$m images are smoothed by a Gaussian kernel of 2.5\arcsec and 12\arcsec, respectively.
    Tails of stripped material extending beyond the optical disc (dotted ellipse) are visible in the H$\alpha$, FIR 250 $\mu$m and HI bands. White arrows indicate the wind direction \citep{vollmer03,lee17}. The red cross in the $g'$-band image of NGC 4330 identifies the background contaminant (see text). North is up; east is to the left.}
    \label{fig:maps}
\end{figure*}
Truncated discs of gas and dust due to environmental effects have already been investigated in the past \citep[e.g.][]{chung09,cortese16,cortese10a,lee17}. Here, we focus on the novelty of the present study, i.e. the identification of more extended H$\alpha$ tails, the detection of diffuse FIR emission of dust tails and their connection with the HI gas component.\\
\indent
In Fig.\ref{fig:maps} we compare the g-band, H$\alpha$, FIR 250 $\mu$m and HI maps for our sample of galaxies, where we smoothed the original H$\alpha$ and FIR images with a Gaussian kernel of 2.5\arcsec and 12\arcsec , respectively, to better show the faint structures. 
In H$\alpha$, FIR, and HI, these galaxies are morphologically asymmetric.
 Furthermore, by comparing the gas and dust distributions with the optical disc extension (white ellipses in Fig.\ref{fig:maps}), besides the well-known truncated discs in H$\alpha$, FIR and HI, a component is visible that extends outside the optical radius. These features are very faint in H$\alpha$ and FIR, reaching the respective survey sensitivity limits in both bands.

\begin{figure*}[h!]
    \centering
    \includegraphics[width=1\textwidth,trim={5.5cm 0cm 7.cm 1cm},clip]{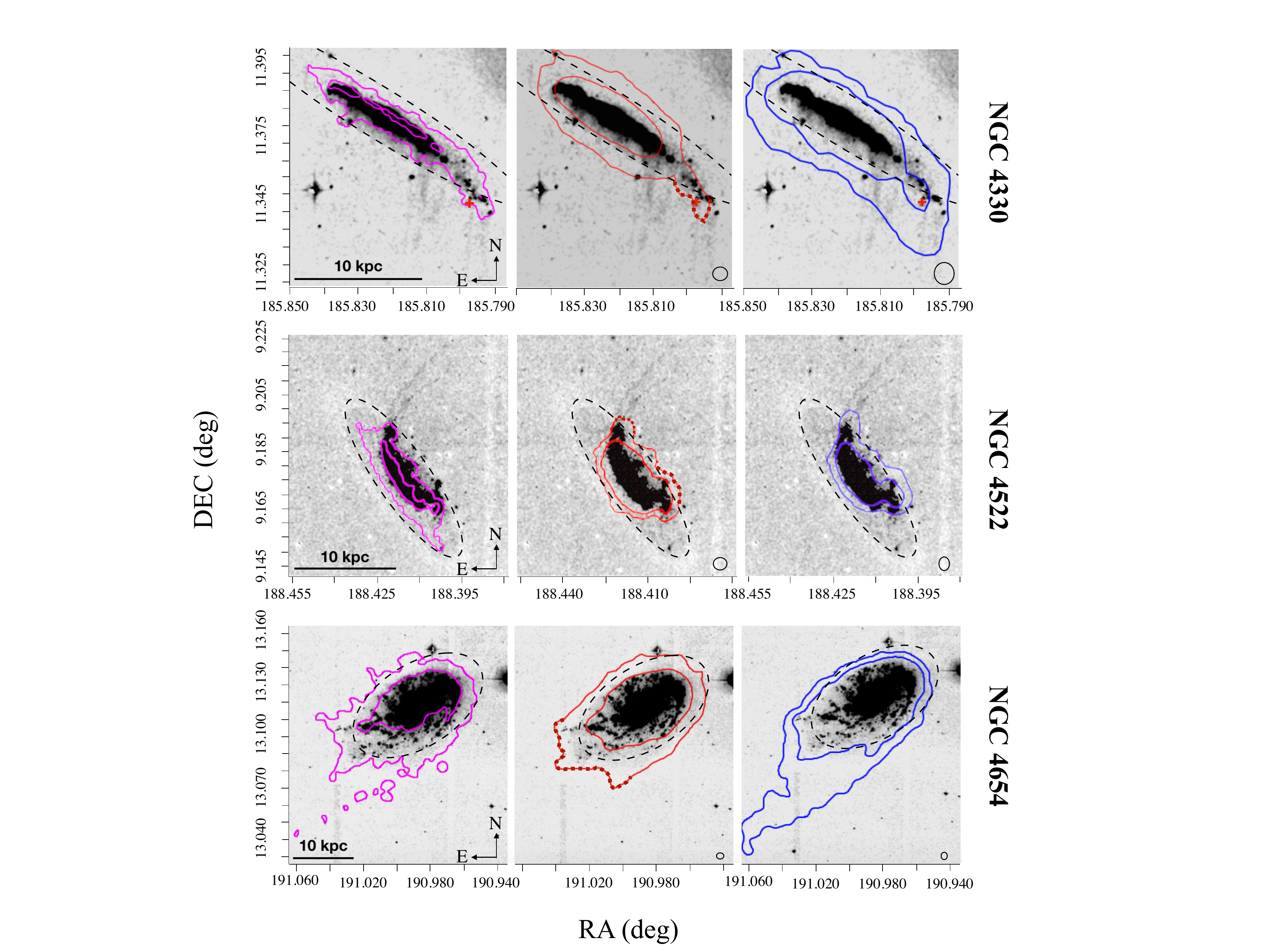}
    \vspace{-0.3cm}
    \caption{Smoothed H$\alpha$ VESTIGE images of NGC 4330 (top), NGC 4522 (center), and NGC 4654 (bottom) compared with the GALEX FUV emission (magenta contours), SPIRE 250 $\mu$m emission (red contours), and VIVA HI emission (blue contours). The faintest emissions from the SPIRE 250 $\mu$m data are at surface brightness levels of 0.6 MJy sr$^{-1}$, while the HI contours levels reach column densities values of $\Sigma_{\mathrm{HI}} = 2 \times 10^{19}\, \mathrm{cm^{-2}}$. Dotted black ellipses trace the extensions of the galaxies optical discs. The regions of the tails outside the optical disc are considered for our photometric analysis (red dotted contours). Linear scales and synthesised beam sizes are shown in the bottom-left and bottom-right corners, respectively.}
    \label{fig:contours}
\end{figure*}
\vspace{-0.4cm}
\noindent
\paragraph{\bf {NGC4330}} The previously detected H$\alpha$ and HI tails bending south in the downstream region of NGC 4330, together with the H$\alpha$ low surface brightness filaments that extend further from the tail to the south \citep[e.g.][]{chung09,fossati18}, are followed by a tail of dust emitting in the FIR that extends out to 6 kpc from the galaxy disc, never detected before.
Superimposed to the H$\alpha$ emission in the tail are regions of recent star formation better seen in Fig.\ref{fig:contours} where the VESTIGE H$\alpha$ image is compared to GUViCS FUV  emission from young stars (magenta contours). Previously identified by several authors in the past \citep{abramson11,boissier12,fossati18}, these features can be explained also by a stripped dust component that cools the gas ablated from the disc and leads to episodes of star formation. The correlation between the ionised gas, dust (red contours), and the HI emission (blue contours) shown in Fig.\ref{fig:contours} may support this hypothesis.
Furthermore, the atomic gas and dust display similar extensions to that of the ionised gas, also covering the region where the low surface brightness H$\alpha$ filaments appear. On the contrary, the FUV emission is limited to the downstream tail and does not cover the regions extending further south, suggesting that we may expect ionised H$\alpha$ emission to have another origin than photo-ionization. In this scenario, the tail hosts massive star formation that ionises the cool gas producing H$\alpha$ emission, while the southern filaments result from the ionisation of stripped atomic gas from thermal conduction or shock-heating due to the interaction with the hot ICM, as recently confirmed in the theoretical work of Vollmer et al. (2020 in press) on NGC 4330 and as found in other ram pressure tails \citep{fossati16} or simulations \citep{tonnesen12} . Finally, the morphology of the filaments, their length, width, and clumpiness may result from the presence of magnetic pressure \citep{fossati18} as also shown by theoretical studies of multi-phase gas stripping \citep[see for example,][]{ruszkowski14,tonnesen14}.

\vspace{-0.5cm}
\noindent
\paragraph{\bf {NGC4522}}
Observed for the first time, two low surface brightness filaments of ionised gas extend by $\sim10$ kpc north from the optically bright galaxy (Fig.\ref{fig:contours}, central raw). The signal, very faint and close to the sensitivity limit of VESTIGE (it has a typical surface-brightness $(1 - 2) \times 10^{-18} \mathrm{erg\, cm^{-2}\, s^{-1}\, arcsec^{-2}}$), defines a very narrow and elongated morphology 
likely suggesting, a dynamically important magnetic field component, as for NGC 4330.
Closer to the optical disc, NGC 4522 is characterised by similar distributions of the ionised and atomic gas, and dust. The H$\alpha$, FIR, and HI images show truncated discs (see Fig.\ref{fig:maps}) above which extraplanar emissions extending towards the north-west and west are clearly visible. FUV emission is present in both extraplanar regions, resembling a front of star-formation triggered by the ram pressure event, and extend also beyond the truncated discs. As argued by \citet{vollmer12}, this is consistent with the galaxy being stripped very recently \citep[several 10 Myr][]{crowl06,vollmer06} and the timescale of gas stripping due to ram pressure is shorter than the timescale of FUV emission ($\sim$100 Myr). 

\vspace{-0.5cm}
\noindent
\paragraph{\bf {NGC4654}} The truncation of the ionised and atomic gas and dust discs on the north-west (Fig.\ref{fig:maps}) are clear signatures of ram pressure, while the long HI tail observed on the south-east is consistent with concurrent acting of ram pressure and tidal stripping. The gravitational interaction with NGC 4639 occurred 500 Myr ago and is also responsible for the disturbed morphology of the stellar component in the south-east \citep{vollmer03}. Focusing on the tails, the stripped component is visible in all bands on the eastern side, with similar spatial distribution in their ionised gas and dust emissions and a more extended atomic gas component. The overlap of the FUV emission in this regions  supports the idea that dust may act as a cooling agent for the neutral gas, provided that the HI density is sufficiently high or that it is present molecular gas, favouring the formation of new stars traced by the Ha and FUV emission. Differences among different components are stretched for the tail extending on the south-east. Here, the HI tail extends out to $\sim30$ kpc from the optical disc, followed by FUV emission but no ionised gas. This sets the time-scale of the star-formation process. In fact, H$\alpha$ traces recent events on timescales of $\sim$ 10 Myr \citep{boselli09,boquien14}, while FUV emission timescales extend to $\sim$ 100 Myr.  
Finally, a tail of dust follows the HI component, although its extension is limited to 9 kpc at the sensitivity of the SPIRE data.\\

The tails we identify in the H$\alpha$, FIR 250 $\mu m$, and HI are all interpreted as ram pressure stripped component. This has been vividly illustrated for the ionised and neutral atomic gas, but still needs to be justified for the dust component. Dust in stripped tails may, in fact, have an in-situ origin and be produced by recent episodes of star formation \citep[e.g.][]{poggianti19}. However, the evidence that i) NGC 4330, NGC 4522, and NGC 4654 are all characterised by truncated dust discs and ii) the resolution of the SPIRE data is not sufficient to identify regions of recent star formation in the FIR (i.e. the dust tails are diffuse at the resolution of the SPIRE data) may support the scenario where the dust we detect in emission is, indeed, stripped. 

Future follow-up studies, that will analyse the dust distribution at higher resolutions and sensitivities, will be instrumental to confirm our conclusion.

 \subsection{Photometry}
 \label{photometry}
As we are interested in bridging the cluster-ICM-galaxy evolution,
the photometric analysis focuses on the portions of the tails extending beyond the optical discs (dotted ellipses in Fig.\ref{fig:maps} and Fig.\ref{fig:contours}).\\
\indent
For each galaxy and photometric band, the analysis is performed on the region within a constant FIR surface brightness level of $\ge 0.6$ MJy sr$^{-1}$ (dark-red dotted lines in Fig. \ref{fig:contours}), which is the noise confusion limit of the SPIRE survey in the 250$\mu$m band.
 As in \citet{fossati18} we compute fluxes as the sum of the pixel values contained within the chosen regions and subtracts a background level that is measured as the median value of 1000 measurements computed in the same size apertures randomly distributed across the image after masking bright stars and the main body of the galaxy. The H$\alpha$ values are corrected for Galactic attenuation \citep[see][for details]{fossati18}. The uncertainty on the fluxes is obtained by standard propagation, quadratically combining the uncertainties of the background (rms-variance of the bootstrap samples) and flux counts. The latter is estimated differently for the different bands: for the Ha images, Poisson statistics is assumed; for the HI images, we used the rms-variance from \citet{chung09} scaled by a factor that takes into account the aperture area; for the FIR images, we used Monte Carlo simulations, generating 100 realisations for each region with fluxes drawn pixel-by-pixel from a Gaussian distribution centred on the observed flux and standard deviation equal to the $Herschel$ SPIRE RMS map.
 The statistics of the simulated data set give us the error associated with the flux measurements.
We note that the adopted technique for the background removal also subtracts the signal coming from the diffuse Virgo IC dust, measured to be constant on much larger scale \citep[$\sim$ 1 deg;][]{longobardi20} than the one we used for the background estimation.
We also emphasise that, because of the relatively large areas we use to define the tail regions, the photometry is not affected by beam smearing even in the HI images characterised by the poorer spatial resolution.\\
\indent
Finally, as already stressed in Sect. \ref{data}, FIR data at wavelengths outside the SPIRE 250 $\mu$m band are too shallow and too poorly resolved to allow for a statistically meaningful measurement of the fluxes. For instance, the same photometric analysis carried out on the PACS images yields flux densities estimates with $\mathrm{S/N} < 1.5$. Therefore, in the following analysis, the results are extracted from the information derived from the 250 $\mu$m images, and the PACS data will only be used to set upper-limits on the temperature of the stripped dust. 
\subsection{FIR Background contamination}

Since the targeted FIR surface brightnesses are close to the confusion limit of the $Herschel$ survey, any possible background contamination must be carefully taken into account. The areas we study are indeed large enough to host background galaxies whose emission contributes to the estimated FIR fluxes.
The spatial distribution of low-luminosity ($z\mathrm{-band\, mag} \ge 20.0$) and compact background sources \citep[extracted from SDSS DR12;][]{alam15} located around our galaxies is homogeneous and well-sampled across the field-of-view of interest. If there is some contamination from these objects not resolved in the FIR images, then their contribution is folded in the background estimates and their uncertainties. 
However, the presence of extended and luminous galaxies may imply a larger contribution that must be taken into account. We identified one bright background object in the stripped tail of NGC 4330 with photometric redshift  $z_{\mathrm{ph}} = 0.14$ and magnitude $z\mathrm{-band\, mag} = 19.6$.

\begin{table*}[t]
\label{galaxy_prop}
    \centering
    \caption{Physical properties of the stripped gas and dust extending beyond the optical disc.}
    \vspace{-0.2cm}
    \begin{tabular}{ccccccccccccc}
    \hline
    \hline
    ID &$\sigma_{100\mu m}$&$\sigma_{160\mu m}$&$S_{250\mu m}$&$S_{HI}$ & ${M_{\mathrm{dust}}}$ & ${M_{\mathrm{HI}}}$\\
     & (mJy)& (mJy)  & (mJy) & (Jy kms$^{-1})$ &($M_{\odot}$) &($M_{\odot}$)\\
     (1)& (2) & (3) & (4) & (5) &(6)& (7) \\
     \hline\\[-5pt]
    NGC4330 &18.4&16.7&18.0 $\pm4.7$& 0.1$\pm0.01$&$3.9\pm1.0$\, e+4&$6.3\pm0.4$\, e+06\\
    NGC4522 &26.0&22.0&29.8$\pm6.1$&0.5$\pm0.03$&$6.4\pm1.3$\, e+4&$3.5\pm0.2$\, e+07\\
    NGC4654 &230.5&183.4&530.1$\pm39.4$&17.7$\pm0.6$&$11.4\pm0.8$\, e+5&$1.1\pm0.1$\, e+09\\
       \hline
    \end{tabular}
    \tablefoot{{\bf Column 1}: Galaxy Name. {\bf Column 2-3}: 1 $\sigma$ uncertainty on the FIR 100 $\mu$m and 160 $\mu$m flux densities within the examined regions. {\bf Column 4-5}: FIR 250 $\mu$m, and HI flux densities within the examined regions. {\bf Column 6}: Dust mass computed from column 3 via Eq.\ref{equation:Dust_mass}, assuming a dust temperature $T = 20\, K$. {\bf Column 7}: Mass of the atomic gas computed from column 4 via Eq.\ref{equation:HI_mass}}.
    \label{tab:physical_p_tail}
\end{table*}

Using the plethora of multi-wavelength data surveying the Virgo region, we estimate the expected flux in the $Herschel$ 250 $\mu m$ band due to background contamination by the broad-band spectral energy distribution (SED) fitting code CIGALE\footnote{http://cigale.lam.fr/} \citep{burgarella05,noll09,boquien19}. Specifically, we use 
a $Salpeter$ initial mass function (IMF) and the SEDs of \citet{bc03} convolved with different exponentially decreasing star formation histories (SFHs). The dust emission is given by the \citet{draine14} models, with a dust attenuation described by a modified \citet{charlot00} attenuation law \citep{buat18}. \
\begin{figure}
    \centering
    \includegraphics[width=0.9\columnwidth,trim={0.3cm 0cm 0cm 0.2cm},clip]{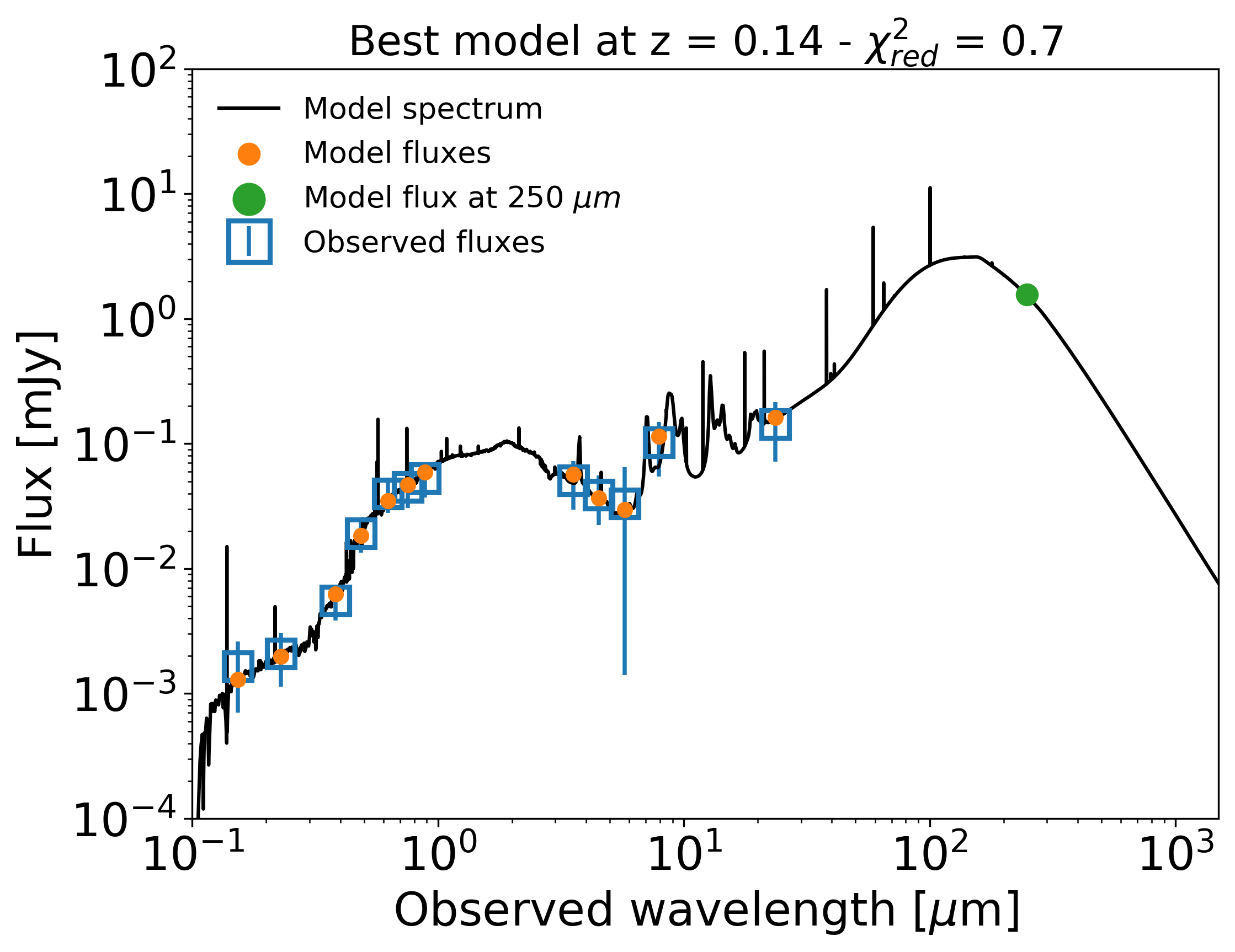}
   
    \caption{ SED of the background galaxy contributing FIR flux in the tail of NGC 4330. Blue squares with errorbars are the observed fluxes with uncertainties. In orange are the modelled fluxes as predicted by the CIGALE fitting model (black line). The green dot identifies the extrapolated flux density at $250\, \mu$m.}
    \label{fig:SED_backgal}
\end{figure}

The observational data used to constrain the SED fitting were limited to the UV - MIR region where the object fluxes are well above the sensitivities and confusion limits of the surveys. In particular, we used GUViCS near- and far-UV data, deep broad-band $u'$ $g'$ $r'$ $i'$ $z'$ data from the NGVS and VESTIGE, and deep $Spitzer$ IRAC and MIPS photometry at 3.6, 4.5, 5.8, 8.0, and 24  $\mu m$ \citep{fazio04,rieke04}.
The CIGALE method is based on an energy balance between the energy absorbed by dust in the UV-optical and the energy re-emitted as IR radiation by the dust itself. We can, therefore, consistently extrapolate the flux redistributed at FIR wavelengths just sampling to the MIR wavelengths \citep{dobbels20}.
The fit gives a 250 $\mu$m flux density (Bayesian estimate) of $\mathrm{F_{CIGALE}}=1.17\pm0.4$ mJy, with an effective reduced chi-square, $\chi^{2}=0.7$. Thus, the background galaxy contributes $\sim10~\%$ to the total flux density in the tail of NGC 4330, that can be subtracted. In Fig.\ref{fig:SED_backgal} we plot the observed and modelled UV to IR SED of the background galaxy to qualitatively show the reliability of the fitting result.
In the Appendix \ref{appendix:CIGALE}, the reader can find a summary of the main photometric properties of the identified background galaxy (Table \ref{tab:back_gal}), the SED fitting results, together with the information on the initial parameters adopted in the fitting (Table \ref{tab:back_gal_CIGALE}).\\

To summarise, on the basis of the photometric analysis of NGC 4330, NGC 4522, and NGC 4654 we measure a statistically significant H$\alpha$ and FIR flux where HI data trace stripped tails of gas \citep{chung09}. Each of the stripped tails then comes with a measurement of the H$\alpha$, HI, and FIR 250 $\mu m$ fluxes in the region extending beyond the optical disc (Table \ref{tab:physical_p_tail} lists the H$\alpha$, HI and FIR fluxes as computed in this work). Thus, we can determine the dust and gas masses for the tails of stripped material and compute the corresponding dust-to-gas ratios. These are the goals of the next session.

\section{Dust-to-gas ratios}
\label{DGR}
\subsection{Dust and gas masses}
For each tail region, the dust mass is obtained via the relation 
\begin{equation}
M_{dust}=\frac{S_{250\mu m}D^{2}}{K_{250\mu m}B_{250\mu m}(T)},
\label{equation:Dust_mass}
\end{equation}
where $S_{250\mu \mathrm{m}}$ is the measured FIR flux density in Jy, $D$ is the distance of the galaxy (assumed to be 16.5 Mpc for all galaxies), $K_{250\mu m} = 4.00\, \mathrm{cm^2\, g^{-1}}$ is the dust grain opacity at $\lambda = 250\mu m$ \citep{draine03}, and $B_{250\mu m}(T)$ is the Planck function for dust at temperature $T_{d}$. It is clear that the value of the dust masses strongly depends on the dust temperature. While dust grains in nearby galaxies radiate with a typical temperature $ T_{d} \sim20\, K$, temperature gradients have also been observed for nearby objects, with a radial decrease towards the outer parts of the galaxies \citep[e.g.][]{galametz12}. We then decided to leave $T_d$ as a free parameter and trace the variation of dust-to-gas ratios for dust temperatures within the range $10\leq T_{d}\le 30\, K$.\\
\indent
Gas masses are computed via the relation
\begin{equation}
M_{gas} = \frac{1}{f_\mathrm{H}}\left(M_{HI} + M_{H_{2}}\right),    
\end{equation}
where $f_{\mathrm{H}} \approx0.74$ is the standard fraction of neutral hydrogen gas with the rest consisting of He and a minor fraction of heavier elements, $M_{H_{2}}$ is the mass of molecular hydrogen,
and $M_{HI} $ is the HI mass in solar units, derived from the HI flux as:
\begin{equation}
M_{HI} = 2.356\times 10^{5} S_{HI} D^{2},
\label{equation:HI_mass}
\end{equation}
with $S_{HI}$ the measured HI flux in Jy kms$^{-1}$ and $D$ as in Eq. \ref{equation:Dust_mass}.\\ 
\indent
\begin{figure*}[t]
    \centering
    \includegraphics[width=18.cm,trim={0cm 6.cm 0.cm 8cm},clip]{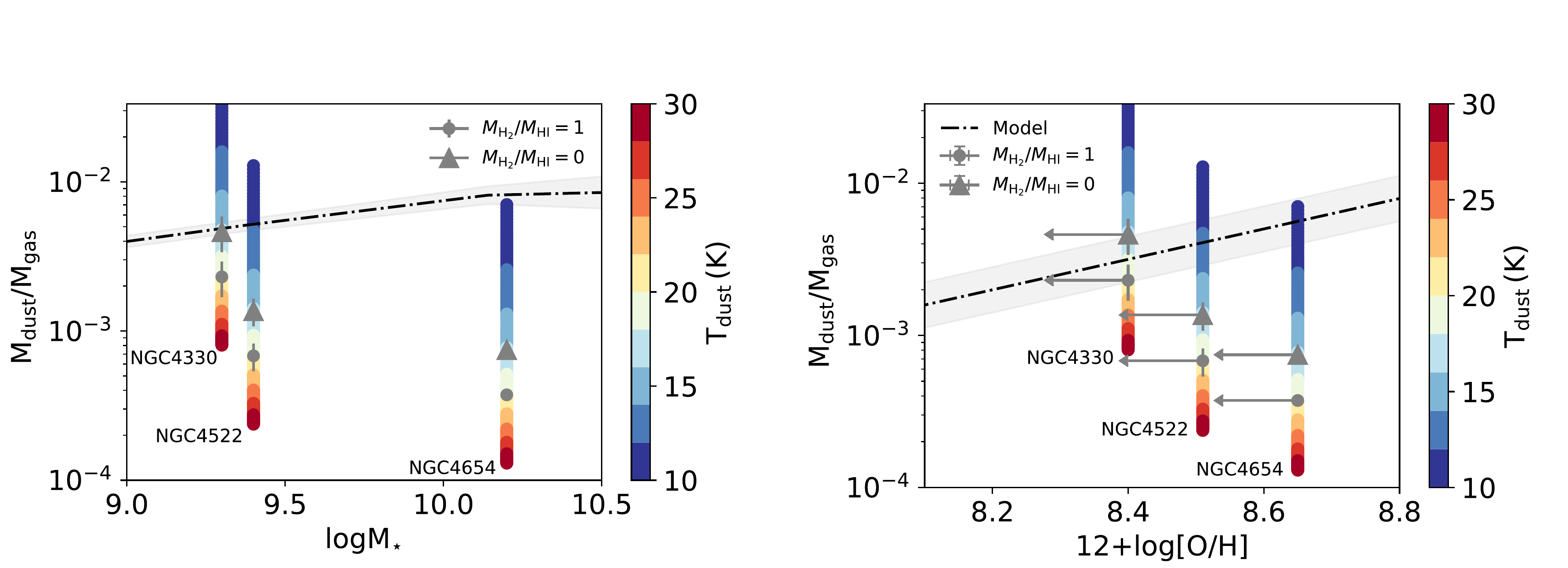}
    \caption{{\bf Left}: Dust-to-gas mass ratios in the stripped tails of NGC 4430, NGC 4522, and NGC 4654 as a function of the
    galaxy stellar mass for several values of dust temperature (color bar). Gray dots (gray triangles) represent the dust-to-gas mass ratio if the stripped dust has temperature $T=20\, K$ and $M_{\mathrm{H_2}}$/$M_\mathrm{HI}$ $=1$ ($M_{\mathrm{H_2}}$/$M_\mathrm{HI}$ $=1$, i.e. no H$_2$ contribution in the tails).  The error bars are dominated by the uncertainties on the FIR flux estimates. The dash-dotted line shows the measured values of the dust-to-gas ratios as a function of mass as measured in Virgo galaxies \citep[][; we plot the case with varying CO-to-H$_2$ conversion factor]{cortese16}. {\bf Right}: As left panel, however this time the dust-to-gas ratios are as function of the mean galaxy metallicity. The arrow plots the relation as a function of the outer galaxy metallicity (see text for further details).  The dash-dotted line traces the expected dust-to-gas ratio metallicity relation for local late type galaxies with the shaded area tracing one standard deviation in a given metallicity bin \citep{moustakas10,hughes13}.}
    \label{fig:ratio}
\end{figure*}

Our estimates of the total gas mass depend on the assumption of the fraction of molecular to neutral atomic gas. Thus, we carry out our analysis under two assumptions: 1) $M_{\mathrm{H_2}}$/$M_\mathrm{HI}$~=~1, i.e. the mean value observed in the local Universe within the main body of the galaxies \citep{boselli14b,cortese16}, and 2) $M_{\mathrm{H_2}}$ = 0, i.e. no contribution by the neutral molecular gas phase. In particular, the second ansatz is based on the observation that the density of the
molecular gas increases towards the galaxy's center \citep[e.g.][]{lee17,casasola17} and that only a few examples have been observed in which both the atomic and molecular gas phases are stripped off during the interaction \citep{vollmer01,vollmer05,vollmer08,jachym17,cramer19,zabel19}. This topic is still under debate and will be one of the main analysis that will be carried out by VERTICO (Virgo Environment Traced in CO survey; Brown et al. in preparation) . Finally, we do not consider the case of  $M_{\mathrm{H_2}}$/$M_\mathrm{HI}$ $>1$. Resulted from an in-situ origin of the molecular gas in the stripped tails as a consequence of the condensation of neutral atomic gas \citep[e.g.][]{verdugo15,moretti18,moretti20}, it would yield bright clouds of molecular gas that have not been detected in previous surveys  of our galaxy sample \citep[only NGC 4522 shows extra-planar emission of molecular gas, with a molecular-to-neutral atomic gas ratio of $\sim 0.5$][]{vollmer12,chung14,lee17}. Thus, gas masses estimated assuming $M_{\mathrm{H_2}}$ $=0$ may represent a lower limit of the real value, the latter likely lying in between the two cases we consider.
From the estimated dust and gas masses (see Table \ref{tab:physical_p_tail} for a list of the measured values) the dust-to-gas ratios in the case of $M_{\mathrm{H_2}}$/$M_\mathrm{HI}$ $=1$ is $2.0\pm0.6\times10^{-3}$, $0.7\pm0.1\times10^{-3}$, and $0.4\pm0.03\times10^{-3}$, for NGC 4330, NGC 4522, and NGC 4654, respectively, that increase by a factor of two when no H$_{2}$ contribution is considered. Their values as a function of the galaxy stellar mass and dust temperature are shown as vertical bars in Fig.\ref{fig:ratio} (left panel). It is clear that by adopting $T_{d} = 20\, K$ (grey dots) the estimates are inconsistent by several standard deviations from the value of 10$^{-2}$ typical of local late-type galaxies \citep[dashed-dotted line; e.g.][]{cortese16}. Even with no contribution from  H$_2$ molecules (grey triangles) the ratios are well below the value of reference for $T_{d} = 20\, K$, and colder dust temperatures would be required.

\subsection{Dust-to-gas ratios vs metallicity relation}
\begin{figure}
    \centering
    \includegraphics[width=9.cm,trim={0.6cm 0.6cm 0.cm 0cm},clip]{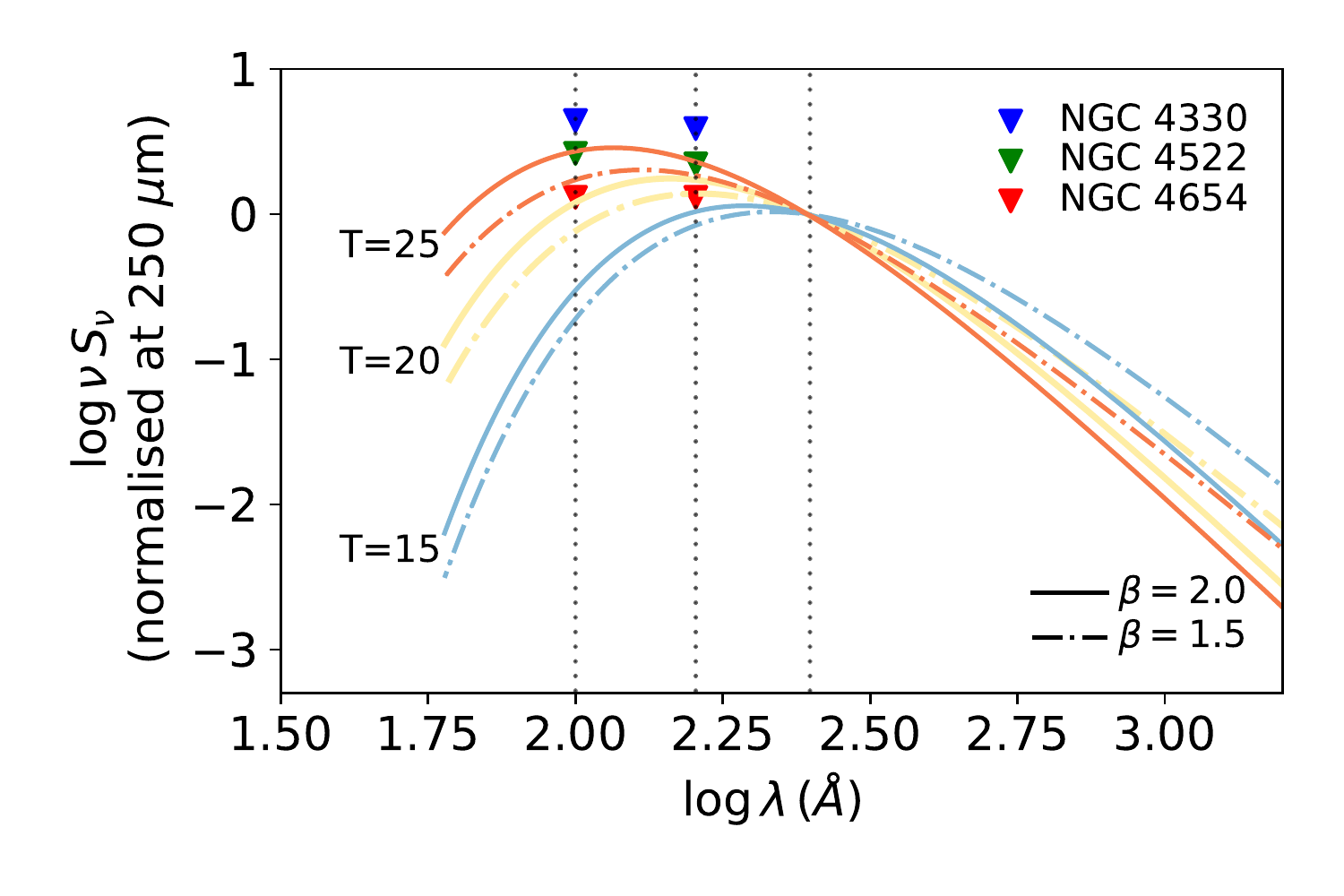}
    \caption{Dust emission simulated using the modified blackbody law for $15\, K$ (blue lines), $20\, K$ (yellow lines), and $25\, K$ (red lines) dust grains. The dust emission assuming a single spectral index of $\beta=2.0$ (continuous lines) and $\beta=1.5$ (dotted-dashed line) are also represented. The curves have been normalized at 250~$\mu$m. The upper limits of the normalised flux densities at 100 $\mu$m and 160 $\mu$m measured in the tails of NGC 4330, NGC 4522 and NGC 4654, are shown as blue, green, and red down pointing triangles, respectively. Dotted vertical lines identify 100 $\mu$m, 160 $\mu$m, and 250$\mu$m.}
    \label{equation:Dust_SED}
\end{figure}

Several studies in the literature have shown that metallicity is the main property of a galaxy driving the observed dust-to-gas ratios. At high metallicities and down to $12 + \log{\mathrm{[O/H]}} \sim 8.2$, the relation between the dust-to-gas mass ratio versus metallicity is well represented by a single power law with a slope of -1 \citep{james02,draine07,galliano08,moustakas10,leroy11,remyruyer14,casasola20}, with a standard deviation of the ratio,  in a given metallicity bin, of $\sim0.15$ dex \citep{moustakas10}.
For our objects the metallicity values can be compiled from \citet{hughes13}, which using drift-scan optical spectroscopy, derived oxygen abundance estimates for a large sample of Virgo late-type galaxies also compiling a stellar-mass metallicity relation.
For NGC 4522 and NGC 4654 they reported a mean metallicity of $\mathrm{12+\log{[O/H]}} = 8.51 \pm 0.48$ and $8.65 \pm 0.07$, respectively. For NGC 4330, which is not in their sample, we estimated the mean metallicity using their mass metallicity relation and find, for a stellar mass of $M_{*} = 10^{9.3} M_{\odot}$, a value of $\mathrm{12+\log[O/H]} = 8.4 \pm 0.1$. These values are compatible with the oxygen abundance estimates derived in other studies \citep[e.g.][]{devis19} and, more in general, they are consistent with the mean metallicity value typical of Scd galaxies \citep{casasola20}. \\
\indent
Fig.\ref{fig:ratio} (right panel) shows the dust-to-gas ratio - metallicity relation for the tails of our galaxies\footnote{Depending on the inclination of the galaxy, thus whether stripping happens mostly edge-on (NGC 4654) or face-on (NGC 4330, NGC 4522), and on the epoch of the stripping process (NGC 4522 is at a later stage with respect to NGC 4330 and NGC 4654), the mixing of the stripped gas component, thus its metallicity, differs from the mean galaxy value. The mean metallicity of NGC 4522, indeed, mostly traces the still existing inner disk only, while NGC 4330 and NGC 4654 both have their discs still existing. These, however, are second order effects well within the uncertainties of the metallicity values we consider.}. Vertical bars trace the range of dust-to-gas ratios obtained assuming different dust temperatures, $10 \le T_{d} \le 30\, K$, (in the case of $M_{\mathrm{H_2}}$/$M_\mathrm{HI}$~=~1) and are plotted against the mean galaxy metallicities. Grey dots (triangles) are representative of the ratios in the stripped tails with a dust component at $T_{d} =20\, K$ accounting (not accounting) for H$_{2}$ molecules. Finally, the arrows point towards the values of the dust-to-gas ratios expected if this relation was plotted as function of the outer galaxy metallicity. It is now well known that late-type galaxies are characterised by negative metallicity gradients towards their outskirts. \citet{moustakas10,magrini11} analysed the variation with radius of oxygen abundances showing that galaxies with similar optical extensions and mean abundances  ($8.4\le \mathrm{\log{[O/H]}} \le 8.7$) as our sample of objects are characterised by negative metallicity gradients with a mean slope of $-0.36\pm0.06$. Such a gradient means that at the optical radius our objects are $\sim0.14$ dex more metal poor than as implied by the mean metallicity value. Fig. \ref{fig:ratio} may suggest a negative trend of the dust-to-gas ratio with metallicity. However, if confirmed on a statistical base, it has important consequences on the physics of the ISM and on the formation of the cluster IC component. We develop these in what follows. \\ 
\indent
There are two considerations we can make from Fig.\ref{fig:ratio} : 1) assuming that the dust-to-gas ratio metallicity relations are to be valid for baryons stripped from the main body of the galaxies, our estimates may follow the theoretical relations (dash-dotted lines) if the stripped dust is cold and emitting at different temperatures in the three galaxy tails, i.e. $T_{d} \lesssim 20\, K$; 2) assuming similar dust temperatures for the different objects, our measurements show a decreasing trend of the dust-to-gas ratio for the stripped material, both with increasing stellar mass and metallicity. Let's comment on these two cases, separately. 
\vspace{-0.5cm}
\noindent
\paragraph*{Dust temperature upper-limits} In Fig. \ref{equation:Dust_SED} we plot the fraction of 100 $\mu$m and 160 $\mu$m to 250~$\mu$m flux densities in the tails of our galaxies (down pointing triangles). As pointed out in Sect. \ref{photometry}, we can only measure upper limits at these wavelengths, computed as $3\times \sigma_{\lambda}$, where $\sigma_{\lambda}$ is the uncertainty associated to the PACS measurements in the tail regions (see Table \ref{galaxy_prop}). These values are compared with the theoretical SED of dust grains emitting as a modified black body spectrum for different $T_{d}$ and dust emissivities $\beta =2.0 - 1.5$ (coloured lines), shown to fit well 
the observed flux density ratios of nearby late-type galaxies \citep{boselli12}.
 Fig. \ref{equation:Dust_SED} implies that the dust in the tails of NGC 4330 and NGC4522 is $T_d \lesssim 25\, K$ and $T_{d} \lesssim 20\, K$ in the stripped component of NGC 4654. This information, once combined with previous studies showing that dust in nearby galaxies never reaches temperature $T_{d} < 15\, K$ \citep[e.g.][]{galametz12} supports the idea that the temperature of dust, stripped from the outskirt of galaxies, is cold, however, not at the temperatures we would require to retrieve the dust-to-gas ratio-metallicity relation representative of the main body of the galaxies.
 
\begin{figure}
    \centering
    \includegraphics[width=9.cm,trim={0.3cm 0.6cm 0.cm 0cm},clip]{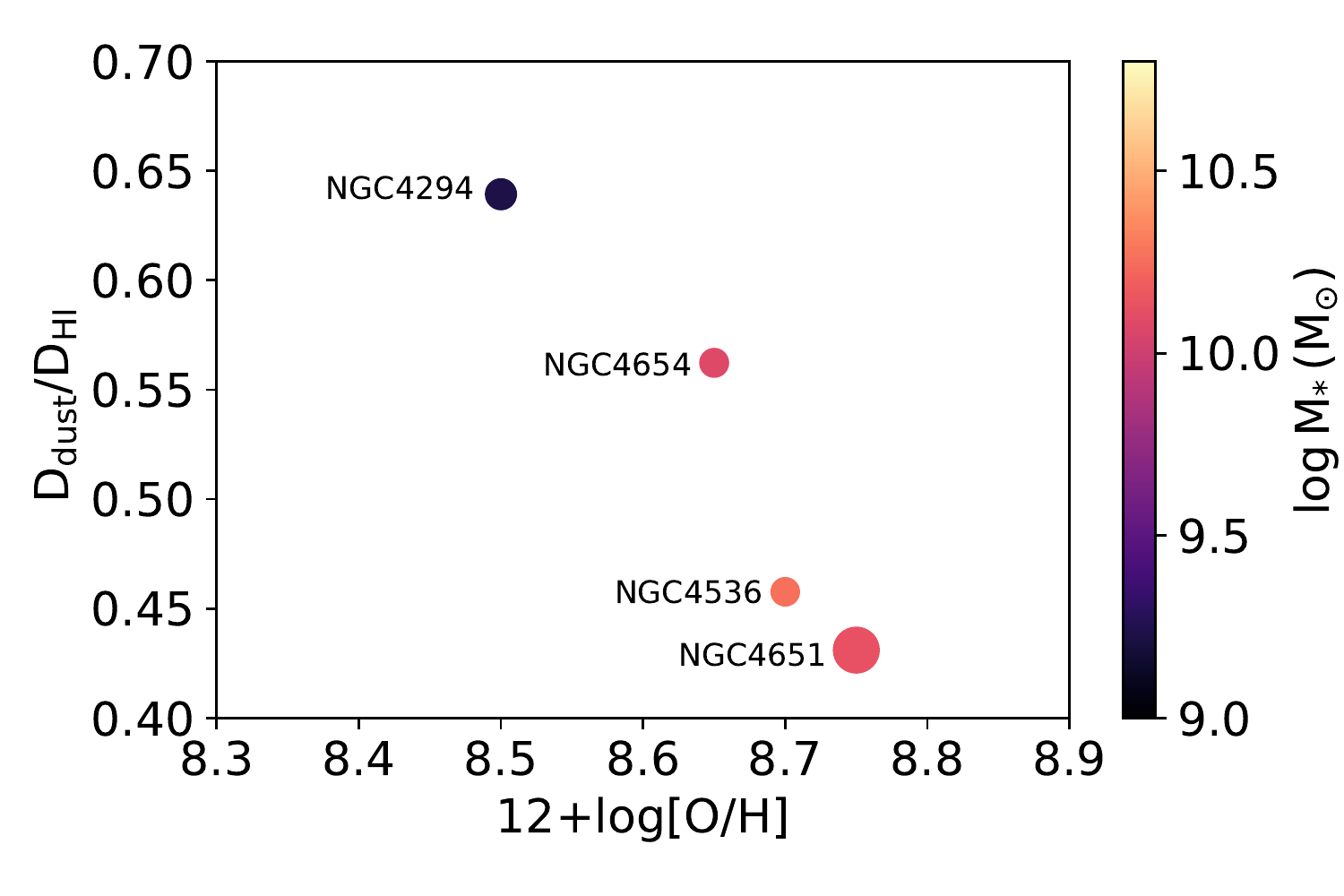}
    \caption{Distribution of the ratios of the dust to HI disc diameters (circles) as a function of the mean galaxy metallicity for non HI-deficient galaxies in Virgo. Objects are colour-coded according to their stellar mass and their sizes is proportional to their HI deficient parameter: the larger the size the smaller the HI-$def$ \citep{hughes13}.}
    \label{Disk_diameter}
\end{figure}

 \vspace{-0.5cm}
\noindent
\paragraph{Dust and gas distribution} To further investigate the anticorrelation we measure for the dust-to-gas-metallicity relation, we focus on how the dust and gas distributions vary within the main body of HI non-deficient galaxies \footnote{The HI$-def$ parameter, measured as the logarithmic difference between the expected and observed HI masses, can be used to quantify the degree of stripping that a galaxy is suffering in the hostile cluster environment.} as a function of the mean galaxy metallicity. To do so, we restrict our analysis to the galaxies of Sb/Sbc/Sc/Scd morphological type, sampled by the $Herschel$ and VIVA surveys, well resolved in the HI images (i.e. with major axis larger than 10 times the beam width) and characterised by $\mathrm{HI}-def < 0.3$ \footnote{The typical dispersion of the HI$-def$ parameter in isolated galaxies is 0.3, thus HI$-def <0.3$ may identify unperturbed systems}. They are NGC 4294, NGC 4536, NGC 4651, and NGC 4654, with mean metallicity $8.5\pm 0.2,\, 8.7\pm0.2,\, 8.75\pm0.07\, \mathrm{and}\, 8.65\pm 0.07$ \citep{hughes13}. 
 Their 250 $\mu$m FIR disc diameters are computed as in \citet{cortese10a}, and taken as the isophotal radii determined at the $6.7 \times 10^{-5}$ Jy arcsec$^{-2}$ surface brightness level, i.e. the average surface brightness observed at the optical radius of non HI-deficient Virgo galaxies. HI discs extensions are the HI isophotal diameters taken at a surface density level of 1 $\mathrm{M_{\odot}\, pc^{-2}}$ \citep{chung09}. In Fig. \ref{Disk_diameter} we show the ratio of the FIR-to-HI disc diameters as a function of the metallicity for the 4 HI non-deficient galaxies. There is a strong correlation between the two variables, implying that in systems not yet 'disturbed' by the cluster environment, the more metal-rich the galaxy, the more centrally concentrated the disc of dust with respect to the disc of HI gas. This finding is supported by the general results that H$_2$ discs are less extended than the neutral atomic component \citep[e.g.][]{leroy08} and agrees well with previous studies showing that dust discs are more centrally concentrated with respect to the HI component in more early-type systems \citep{bendo03, thomas04}. Although the result would benefit from a larger statistics, we can understand from physical reasons that the values and the decreasing trend in dust-to-gas ratio in the stripped tails we measure may be the consequence of a dust component more strongly bound to the gravitational potential well of the galaxy with respect to the HI gas will be less easily removed in any kind of interaction, leading to a smaller dust-to-gas ratio in the tails of more metal rich systems.\\
 \indent
In Sect.\ref{dust_stripping} we also discuss whether the measured trend in dust-to-gas ratio may come from  different sputtering times characterising the dust during the stripping process.

\section{Discussions}
\subsection{Multi-phase gas coexistence in stripped tails}
In Sect.\ref{stripped_tails} we have shown evidence of multi-phase gas stripping showing the coexistence of H$\alpha$ and HI gas in the stripped tail of NGC 4330, NGC 4522, and NGC 4654, increasing the number of known Virgo galaxies with tails in both gas phases (the already known ones being NGC 4522, NGC 4438, and NGC 4330; \citet{kenney04, oosterloo05,fossati18}). 
Several explanations have been given for the coexistence of both phases, mostly related with the interpretation of how the H$\alpha$ emission is produced. In fact, filaments of ionised gas can be explained being i) gas shock-ionised by the galaxy's AGN, ii) gas heated by thermal conduction from the ICM, iii) gas heated by turbulent shocks, 
or iv) gas ionised due to the presence of recent star-formation.\\ 
\indent
A detailed analysis of NGC 4330, carried out as part of the VESTIGE initiative by \citet{fossati18}, showed that the H$\alpha$ emission in the stripped region is partially due to photoionisation by UV radiation coming from a generation of young stars formed in compact regions, and partially due to ionisation as a consequence of the interaction with the hot ICM. A similar interpretation can describe the cases of NGC 4522 and NGC 4654. In the areas where there are compact regions of H$\alpha$ in the tails of NGC 4522, GALEX data shows that there is FUV emission closely following the ionised gas distribution. Further north, instead, where VESTIGE data have revealed the presence of filaments of ionised gas, we may speculate that there should be stripped HI gas, whose ionisation yields to diffuse H$\alpha$ emission although at a lower surface brightness levels than the VIVA survey sensitivity. FUV emission, well correlated with the HI surface density, overlaps as well with the less extended tail of ionised gas emission in the western region of NGC 4654. As emphasised in Sect. \ref{stripped_tails}, the more extended and diffuse HI tail in the south-east region, also visible in the FUV, is not detected in H$\alpha$ tracing the different timescales characterising the H$\alpha$ and FUV emission (roughly a factor 10 shorter for the H$\alpha$ emission). 
Thus, when there is a lack of new stars formed in compact regions, the presence of tails extending beyond the galaxy stellar disc (in our case $\sim 10$ kpc away) leads us to speculate that the gas is mainly excited by mechanisms other than photoionization. These could be shocks in the turbulent gas, magneto-hydrodynamic waves, and heat conduction. In fact, the presence of thin filamentary structures that can be observed along the tails, suggests that magnetic fields might play an important role \citep[see][for a detailed discussion]{boselli16}.

We might also wonder whether, together with the ionised and atomic gas phases, additional gas components emitting at different wavelength are also stripped.
 Studies of CO in NGC 4330, NGC 4522, and NGC 4654 have not revealed stripped tails of H$_2$ gas, even though the inner molecular gas distributions are affected by the strong ICM pressure \citep{vollmer12,chung14,lee17}. Outside our sample, molecular gas in the tail of stripped material has been detected in the Norma galaxy ESO 137-001 for which the authors suggest that the most likely scenario of H$_2$ formation combines the removal of molecular gas that survived ram pressure stripping and the in-situ formation out of stripped atomic gas \citep{jachym17}. The latter scenario is also used to describe the presence of not negligible molecular gas in the tails of several stripped galaxies at redshift $z\sim 0.05$ \citep[][]{moretti18,moretti20}.\\
 \indent
 Therefore, within a typical environment with characteristics similar to those encountered in the Virgo cluster ($\mathrm{n_{e} \sim 10^{-3}\, cm^{-3}},\, T\sim10^{7}\, K$, ) the gas in the stripped component can either be hot and low-density, or cold ($T\sim 10-100\, K$) and high-density  ($\mathrm{n_{e} \sim 100\times 10^{-3}\, cm^{-3}}$) \citep[e.g.][]{vollmer01}. When the density of the gas is sufficiently high, new stars can be formed. On the other hand, if there is a dust component in the stripped tails, as we find in this work, it may act as a cooling agent for the gas \citep[e.g.][]{hollenbach71}, favoring the formation of giant molecular clouds and star forming regions within the stripped material. This, may foster new episodes of star formation and contribute to the star-forming regions found to inhabit ram pressure stripped tails. To which extent star formation is a common phenomenon in ram-pressure stripped tails is still under debate \citep{hester10,fumagalli11,kenney14,boissier12,boselli16,fossati16,gullieuszik17, poggianti19, bianconi20}. One hint may come from the study by \citet{steyrleithner20} in which they found by numerical simulations of ram-pressure stripped dwarf galaxies that obviously high relative velocities with respect to the ICM are necessary to dissolve sufficiently massive clouds which remain gravitationally bound and are capable
to cool, collapse, and form star clusters.
 



\subsection{Dust stripping in Virgo}
\subsubsection{Stripped dust detected in emission}
\label{dust_stripping}
Since the dust in galaxies is associated with the gaseous component of the ISM, it is generally expected that when the gas is stripped part of the dust is removed as well. Supporting this idea are the studies of the gas and dust content in cluster members showing that systems approaching regions of high density are found to be redder and gas, dust deficient with respect to the population of galaxies in the field \citep{boselli06a,gavazzi10,cortese12,boselli14a}.\\
\indent
In this work we measure direct emission from cold dust that overlaps with the tails of ionised H$\alpha$ and HI in NGC 4330, NGC 4522, and NGC 4654. From the analysis of $Herschel$ FIR 250 $\mu$m and VIVA HI data, in Sect. \ref{DGR} we also estimate that the dust is widespread within the stripped tails with a dust-to-gas ratio, $\mathrm{M_{dust}/M_{gas}} \sim 10^{-3}$,  significantly lower than the value found from the analysis of the main body of nearby galaxies, when we assume a dust component emitting at $T_{d} = 20\, K$. Temperature upper-limits, set with the information retrieved from the 100 and 160 $\mu$m bands, confirm that the stripped dust must be cold (the strongest limit is for NGC 4654 with $T_{d} \lesssim 20\, k$), yet not cold enough to make us retrieve the well known relation between dust-to-gas mass ratio and metallicity. Using $Herschel$ data at far-infrared and submillimetre wavelengths, \citet{galametz12} analysed a subsample of the KINGFISH galaxies \citep{kennicutt11} and for these found the dust temperature varies spatially with a radial decrease towards the outskirts reaching minimum values of $T_{d}\sim15\, K$. 
Thus, cold dust in stripped tails fits well within the scenario that galaxies interacting with the cluster environment are stripped of their outermost regions but it is yet not enough to explain the low dust-to-gas ratios we measure. \\
\indent
By analysing trends in the dust-to-HI disc extension with metallicity for galaxies not yet 'disturbed' by the cluster environment (i.e. non HI defiecient), we find that decreasing with metallicity is the relative extension of dust-to-HI discs.  This result implies a lower dust-to-gas ratio in the stripped material of more metal-rich galaxies simply because these systems have more centrally concentrated dust discs that result less severely affected by the outside-in stripping of the ISM.\\
\indent
Nevertheless, it is worth noting that the retrieved estimates could be explained in terms of different dust emissivities. Lower values may imply larger dust masses, hence higher dust-to-gas ratios. In support of this argument is the study by \citet{bianchi19}. Within the DustPedia project \citep{davies17} these authors analysed the variation of dust emissivity for a sample of 204 spiral galaxies and observed that there is a variation in dust emission properties for spirals of earlier type and higher metallicity, on average lower than the one measured for the Milky Way. Moreover, \citet{ysard19}  showed that dust masses may vary of a factor up to 20 by assuming different grain properties (e.g. chemical composition). Finally, dust is expected to survive sputtering by the harsh X-ray emitting gas on a typical timescale of $1.4$ Myr \citep{draine79}. However, we may wonder whether dust grains can be more efficiently sputtered in the stripping process, leading to the observed low dust-to-gas ratios. Recent cosmological hydrodynamical simulations of cluster evolution that include dust production, growth, supernova-shock-driven destruction, ion-collision-driven thermal sputtering, and high-temperature dust cooling through far-infrared re-radiation of collisionally deposited electron energies, have shown that
the typical thermal sputtering time-scales can be as short as 10 Myr \citep{vogelsberger19}. However, such low time-scales are reached where the simulated ICM gas is hotter ($4-7\times$) and denser ($10-30\times$) than the cluster regions our galaxies reside in \citep[e.g.][]{urban11}. In addition, we also notice that shocks are unlikely to increase the sputtering rate of dust grains as shown by \citet{popescu00} who measured that already at 0.3 Mpc from the cluster centre, a sudden interaction of a Virgo-like galaxy with the cluster ICM would generally drive a shock wave with a typical speed well below the minimum value of $\sim 100\, \mathrm{km\, s^{-1}}$ needed for sputtering.



\subsubsection{ICM metal enrichment by dust stripping}
 In galaxy clusters a number of processes can remove metals from the hosting galaxies and transport them into the intra-cluster space. Previous works that studied the enrichment of the ICM have focused mainly on three mechanisms whereby metals could be removed from a galaxy: 1) metal-enriched gas can be removed as a consequence of gravitational interaction and/or ram pressure stripping; 2) gas can escape the galaxy potential well as the result of the energy input from SN explosions; 3) dust can be ejected in the ICM if the radiation pressure on dust grains due to stellar light may exceed the gravitational force of the matter \citep[e.g.][]{aguirre04}. Also, at the cluster central region AGN wind-driven mass-loss may be responsible for the ICM metal-enrichment \citep[e.g][]{qiu20}.\\
\noindent
The evidence we present in this study of dust stripped due to ram pressure, once combined with previous results supporting dust stripping by the tidal interactions \citep[e.g.][]{cortese10b,gomez10},
leaves little doubt that dust stripping is an additional mechanism for injecting dust grains into the ICM, thus contributing to its metal enrichment. Within the virial radius of Virgo and in the last 125 Myr, we expect that ram pressure stripping has contributed $M_{d} \ge 0.4 \times 10^{9} \mathrm{M_{\odot}}$ to the dust (see next section). \citet{devis19} showed that galaxies with metallicity above $\mathrm{12+\log{[O/H]}} = 8.2$ are characterised by a roughly constant dust-to-metal ratio $M_{d}/M_{Z} \sim 0.2$. Thus, within the central region of Virgo, ram pressure may contribute $M_{Z} \ge 0.8 \times 10^{8} \mathrm{M_{\odot}}$ to the metals mass.
This is consistent with numerical simulations which predict that ram pressure alone can already contribute $\sim 10~\%$ to the enrichment of the ICM in clusters \citep{domainko06}. 

\subsection{The building up of the Virgo intra-cluster component}
\begin{figure*}
    \centering
    \includegraphics[width=0.9\textwidth,trim={5cm 0.2cm 5cm 0.2cm},clip]{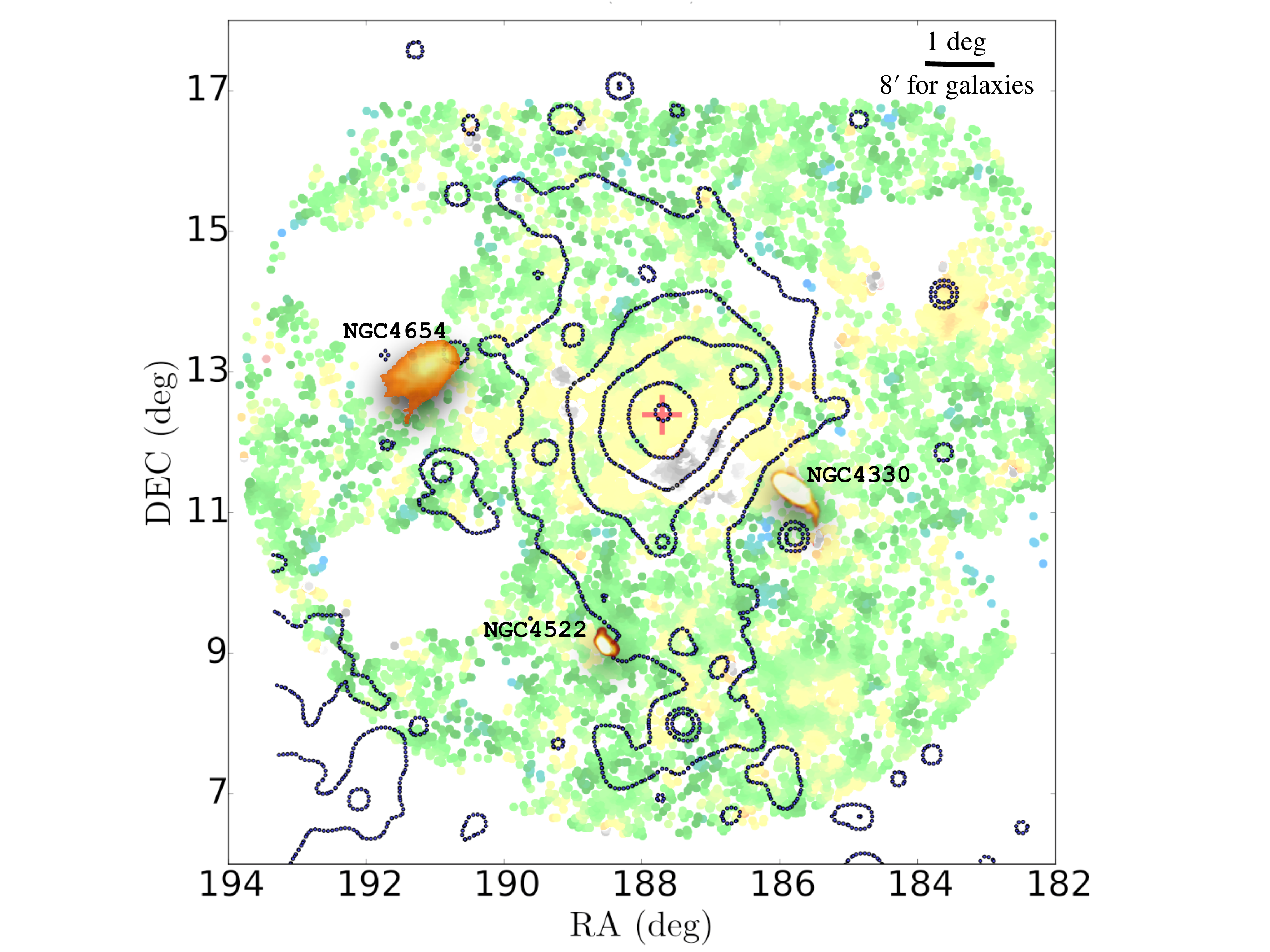}
    \caption{$Herschel$ FIR 250 $\mu$m  images of NGC 4330, NGC 4522, and NGC 4654 (orange) overlaid on the mean extinction map of the Virgo intra-cluster dust (green and yellow trace mean values $<E(B-V)>=0$ and $<E(B-V)>=0.07$, respectively) by \citet[][ here the regions in space contaminated by background clusters have been saturated in grey.]{longobardi20}. The three galaxies are magnified by a factor $\sim 13$ to show the details of the FIR emission. The extinction map shows that the intra-cluster dust is present within $\sim$1.2 Mpc (0.7 virial radii) around the dynamical centre of the cluster (sub-cluster A), as identified by the X-rays emission \citep[Rosat contours from][]{boeringher94}.}
    \label{fig:ICD}
\end{figure*}

 In a hierarchical Universe the presence of an IC component (ICC), i.e. baryons free floating in the cluster potential, is a natural result of the fact that young concentrations of mass are actively in the process of forming.
 The Virgo IC stars, or IC light, have long been studied through their optical photometric and kinematic properties, showing that galaxy interactions, as well as tidal interactions between galaxies and the cluster potential, play an important role in the production of the ICL
 \citep[e.g.][]{arnaboldi02,aguerri05,doherty09,longobardi13,durrell14,longobardi15,hartke17,longobardi18a,longobardi18b,mihos17}. However, if ram-pressure actively strips gas and dust from the galaxies moving through the cluster \citep[e.g][]{cortese10b,cortese12,verdugo15,gullieuszik20,longobardi20},
 it may become an additional key process that contributes to the building up of the IC component. In a recent work, \citet{longobardi20} first detected a diffuse dust component in the intracluster medium of Virgo with mass $2.5\pm0.2 \times 10^{9}\, \mathrm{M_{\odot}}$ and consistent with being the result of stripping phenomena happening out to a scale of 0.6 virial radii.
 We might then wonder whether the results presented in this work support this evidence.
 
 NGC 4330, NGC 4522, and NGC 4654 lie at 0.6, 0.9, and 0.9 Mpc from the centre of the cluster, respectively. Despite their significant distance from the densest and hottest region of the Virgo ICM, ram pressure is actively moving dust and gas outside the main body of the galaxies. Previous works have shown that, when no additional processes such as tidal stripping are in place (like for NGC 4654), this is yet possible as a consequence of the dynamical youth of the Virgo cluster causing local enhancements of the ram pressure due to bulk motions, clumpy density distributions, and variations in the temperature of the ICM gas \citep{kenney04, tonnesen07}. This allows stripping to be effective even in the outskirt of the cluster and supports the finding of a diffuse intra-cluster dust component out to large distances from the centre of the cluster (see Fig.\ref{fig:ICD}).\\
 \indent
 Furthermore, the stripped tails of  NGC 4330, NGC 4522, and NGC 4654 extend beyond the optical discs of the galaxies and are going to be removed from the spirals to fill the intra-cluster space. More specifically, simulations of ram pressure stripping have shown that the ICM-galaxy interaction is marked by different phases \citep{roediger05}. An initial phase, in which the outer part of the gas disc is displaced but only partially unbound, is followed by a second phase during which only a small fraction (about 10~\% of the initial gas mass) of the stripped gas falls back while the remaining fraction is unbound to the galactic potential. For galaxies moving through the Virgo cluster, feeling a ram pressure $p_{\mathrm{ICM}}\ge 1000\,  \mathrm{cm^{-3}km^{2}s^{-2}}$, the second phase is expected to start after 20 Myr, lasting for 200 Myr. If we then assume that ram pressure equally affects all the ISM components, we can expect the stripped dust to be unbound on similar timescales and to become intra-cluster component before it is destroyed by sputtering phenomena in the harsh X-ray environment after 140 Myr \citep{draine79}.
 
 To estimate the total amount of dust in the core of Virgo due to ram-pressure we compute the total gas mass expected to be lost by galaxies undergoing ram-pressure stripping, re-scaling this value by the dust-to-gas mass ratio we found to characterise the stripped ram pressure tails. 
 
 Within the virial radius, Virgo contains 48 late-type galaxies with stellar masses $\mathrm{M_{*} \ge 10^{9}\, \mathrm{M_{\odot}}}$ \citep{boselli14a} and a large fraction of them ($\sim 92~\%$) has measured HI masses and HI deficiency parameters. 
 We can then estimate the amount of atomic and total gas lost being  $\mathrm{M_{HI}} = 0.1\times 10^{12}\, \mathrm{M_{\odot}}$, or $\mathrm{M_{gas}} = 0.3\times 10^{12}\, \mathrm{M_{\odot}}$ \footnote{For the estimation of the total gas mass lost we are considering $M_{\mathrm{H_2}}$/$M_\mathrm{HI}$ $=1$ and a neutral hydrogen gas fraction of 0.74}. For the remaining fraction of galaxies with no HI data ($\sim8~\%$) we can compute a similar estimate by converting the object stellar masses in HI masses using the scaling relation presented in \citet{boselli14b} and assuming an HI deficiency parameter equal to $HI-def=0.92$, i.e. the mean value measured in Virgo A. An additional $\mathrm{M_{HI}} = 0.02\times 10^{12}\, \mathrm{M_{\odot}}$ is lost resulting in a total gas component dispersed within the cluster virial radius of $0.05\times 10^{12}\, \mathrm{M_{\odot}}$. In Sect. \ref{DGR} we have shown that the dust is widespread within the stripped tails with a typical dust-to-gas ratio of $\sim 10^{-3}$ (here we are considering the mean value of the measured dust-to-gas ratios in the case of $M_{\mathrm{H_2}}$/$M_\mathrm{HI}$ $=1$), implying that $\sim 0.4\times 10^{9}\, \mathrm{M_{\odot}}$ of dust is stripped from the main body of the galaxies. If we now consider that the current Virgo accretion rate, for galaxies with $\mathrm{M_{*}} > 10^{9}\, \mathrm{M_{\odot}}$, 
 is $\sim400\,  \mathrm{Gal\, Gyr^{-1}}$ \citep{boselli08,gavazzi13}, we expect $\sim50$ galaxies in 125 Myr,  roughly consistent with the dust survival time before sputtering happens. Thus ram pressure contributes to the diffuse ICD with a dust injection rate of $\sim 3.2 \mathrm{M_{\odot}\, yr^{-1}}$.  The mass estimate we computed above is a lower limit of the amount of dust expected in the IC space by stripping phenomena because i) we considered a mean value of the dust-to-gas ratio to convert gas masses in dust masses, while it is likely that different values must be considered for systems of different mass and metallicities; ii) we have not considered low mass galaxies ($\mathrm{M_{*}} < 10^{9}\, \mathrm{M_{\odot}}$), for which dust stripping is more severe; iii) we have not considered the contribution from dust formed in-situ in ram-pressure stripped tails; iv) 
 several processes may contribute to dust injection in the IC space, among which there are winds of red giant and supergiant IC stars that contribute to the ICD production, even though their contribution is small (dust injection rate of 0.17 $M_\odot \, \mathrm{yr^{-1}}$ \citet{popescu00}).\\
 \noindent
 Therefore, our results set the bases for follow-up studies that will benefit from the higher sensitivity of future missions and map the variation of the dust-to-gas ratio over a larger range of stellar masses.





\section{Summary and Conclusions}
In this work we link the cluster-ICM-galaxy evolution at small scales analysing the relative distribution and fraction in mass of the different baryonic components in the stripped ISM of galaxies that will build up the cluster intra-cluster component. Based on the synergy between VESTIGE H$\alpha$, $Herschel$ FIR 250 $\mu$m, and VIVA HI data we have measured the first direct emission from stripped dust that follows the tails of ionised and atomic gas component. NGC 4330, NGC 4522, and NGC 4654 are the three Virgo galaxies target of our study. They are systems with stellar masses in the range $10^{9}\lesssim \mathrm{M_{*}} \lesssim 10^{10}\, \mathrm{M
_{\odot}}$ , all lying within the cluster virial radius. As such their masses are consistent with the mass of the progenitors of the Virgo ICC \citep[e.g.][]{longobardi18b,pillepich18}, and their projected distances relate them with the diffuse ICD emission, measured to be present within 4 degree from the cluster centre \citep{longobardi20}. More specifically, our results show the following:

\begin{itemize}
    \item NGC 4330, NGC 4522, and NGC 4654 are Virgo galaxies with multi-phase components in their tails of stripped material where cool dust and atomic HI gas overlap with an ionised, hotter, H$\alpha$ emission. For NGC 4330 and NGC 4522 the ionised emission also shows the presence of narrow filaments consistent, given their morphology, with being supported by magnetic pressure. 
    {\b fFuture studies will address the question as to whether the presence of dust in the tails acts as gas cooler and catalyzer of molecular clouds, giving us insights on the detection of star formation processes in the stripped component.}
   
    \item We find that the stripped component is characterised by  dust-to-gas mass ratios of $\sim10^{-3}$, i.e. up to a factor 15 lower than the values measured for the main body of the galaxies. Our analysis also suggests that the metallicity dependent trend is also opposite: the lowest values are associated to the most metal-rich systems.
    \item The stripped dust must be cold, $T_{d} \lesssim 25\, K$ for NGC 4330 and NGC 4522, and even colder, $T_{d} \lesssim 20\, K$, for NGC 4654. 
    Furthermore, 
    the low values and negative trend we measure for the dust-to-gas ratio metallicity relation can be explained with a dust component that is more centrally concentrated in more metal-rich systems. These results well fit within the general picture that galaxies are mainly stripped of their outermost regions, also known to to host colder dust with respect to the hotter, $T_{d}\sim20\, K$, component in the inner regions. However, further observations at different wavelengths and at higher resolutions are necessary to properly sample the FIR and submillimeter regime of the SEDs in the stripped tails and better characterise the dust physical properties like dust temperature and emissivity, the latter also responsible for variation of the dust-to-gas ratios.
    \item The detection of stripped tails of dust within 3 degrees from the cluster centre, is consistent with the recent finding of diffuse dust in the ICM of the Virgo cluster detected out to 1 Mpc. Moreover, under the assumption that ram pressure acts similarly on the dust and gas components, baryons will start being unbound from the galaxy few tens of Myr after the starting of the stripping process. This time scale makes it possible for the dust to become a diffuse component in the ICM of Virgo before being destroyed as consequence of sputtering phenomena, supporting results from simulations showing that the outer regions of dynamically young clusters like Virgo, have as source of intra-cluster grains from ongoing accretion of freshly infalling spiral galaxies \citep{popescu00}.
\item Ram pressure, together with tidal interactions, is a key ingredient for the building up of the Virgo intra-cluster component and highlights dust stripping by tidal interactions and ram pressure as one further mechanism for injecting dust grains into the ICM, thus contributing to its metal enrichment with a dust injection rate of $\sim 3.2 \mathrm{M_{\odot}\, yr^{-1}}$. 
    
\end{itemize}

\begin{acknowledgements}
We thank L. Cortese for providing us with the 250 $\mu$m isophotal radii of a sample of Virgo galaxies. We are grateful to the CFHT team who assisted us in the observations: T. Burdullis, D. Devost, B. Mahoney, N. Manset, A. Petric, S. Prunet, K. Withington.  AL and AN have received funding from the French Centre National d’Etudes Spatiales (CNES), MF from the European Research Council (ERC) (grant agreement No 757535).
This research has made use of data from HRS project. HRS is a Herschel Key Programme utilising Guaranteed Time from the SPIRE instrument team, ESAC scientists and a mission scientist.
The HRS data was accessed through the Herschel Database in Marseille (HeDaM - http://hedam.lam.fr) operated by CeSAM and hosted by the Laboratoire d'Astrophysique de Marseille.
\end{acknowledgements}

\begin{appendix}
\section{CIGALE fitting}
\label{appendix:CIGALE}
This appendix provides additional information on the FIR back ground contaminant in the tail of NGC 4330. In Table \ref{tab:back_gal} we list the main photometric properties of the identified background galaxy and the SED fitting results. Table 
lists the initial parameters adopted in the fitting.

\begin{table}[h!]
    \centering
    \vspace{-0.2cm}
    \caption{Summary of the measured and derived quantities for the background galaxy in the tail of NGC 4330.}
    \resizebox{0.25\textwidth}{!}{
    \begin{tabular}{cc}
    \hline\hline
    &Galaxy 1\\[3pt]
    \hline
    \\[-5pt]
    RA (deg)&185.797\\
    DEC (deg)&11.342\\
    z&0.14\\
    \hline
    \hline 
    \\[-5pt]
    GALEX&\\ [3pt]
      FUV (mag)& $23.3\pm0.2$ \\
      NUV (mag)  & $23.1\pm0.1$\\
      \hline\hline
      \\[-5pt]
      NGVS&\\[3pt]
        $u$ (mag)&$22.05\pm0.01$ \\
        $g$ (mag)&$20.7\pm0.003$ \\
        $r$ (mag)&$19.8\pm0.003$ \\
        $i$ (mag) &$19.8\pm0.004$\\
        $z$ (mag) &$19.6\pm0.006$ \\
        \hline\hline
        \\[-5pt]
        $Spitzer$&\\[3pt]
        3.6 $\mu$m (mJy)& $0.051\pm0.005$ \\
        4.5 $\mu$m (mJy)& $0.039\pm0.004$ \\
        5.8 $\mu$m (mJy)& $0.033\pm0.010$\\
        8.0 $\mu$m (mJy)& $0.102\pm0.012$ \\
        24 $\mu$m (mJy)& $0.143\pm0.019$ \\
        \hline\hline
        \\[-5pt]
        CIGALE&\\[3pt]
        250 $\mu$m (mJy)&$3.9\pm0.4$\\[2pt]
       \hline
    \end{tabular}}
    \vspace{+0.2cm}
    \tablefoot{{\bf Row 1\&2}: J2000 coordinates. {bf Row 3}: photometric redshift from SDSS DR12. {\bf Row 4\&5}: GALEX near and far-UV magnitudes. {\bf Row 6-10}: SDSS optical $u,g,r,i$, and $z$ magnitudes. {\bf Row 11-14}: $Spitzer$ IRAC and MIPS (Row 14) near-infrared fluxes. { \bf Row 15}: Expected flux at 250 $\mu$m as returned by CIGALE SED fitting.}
    \label{tab:back_gal}
\end{table}

\begin{table}[h!]
    \centering
    \vspace{-0.2cm}
    \caption{Input parameters used for the CIGALE SED fitting of the background galaxy in the tail of NGC 4330.}
    \resizebox{1.\columnwidth}{!}{
    \begin{tabular}{cc}
    \hline\hline
    Parameter & Value\\[3pt]
    \hline
    \\[-5pt]
    Pop. synth. mod.& \citet{bc03}\\
    Dust model &\citet{draine14}\\
    IMF & Salpeter\\
    Metallicity& 0.0001, 0.0004, 0.004, {\bf 0.008}, 0.02, 0.05 \\
    $E(B-V)_{\mathrm{young}}$&0.1, {\bf 0.3}, 0.5, 0.7, 1.0\\
    $E(B-V)_{\mathrm{old}}$& {\bf 0.44}\\
    $Q_{\mathrm{PAH}}$&1.12, 2.5, 3.19, {\bf 4.58}, 5.95\\
    $\alpha$&{\bf 2.0}\\
    $\gamma$&{\bf 0.02}\\
       \hline
    \end{tabular}}
    \vspace{+0.2cm}
    \tablefoot{ {\bf Row~1}:~Stellar population model {\bf Row~2}:~Dust emission model. {\bf Row~3}:~Adopted initial mass function. {\bf Row~4}:~Metallicity. {\bf Row~5-6}:~V-band attenuation in the interstellar medium. { \bf Row~7-9}:~Dust model input parameters Mass fraction of PAH. Parameters for the final model used to trace the variation of the flux densities in the background galaxies are given in boldface.}
    \label{tab:back_gal_CIGALE}
\end{table}

\end{appendix}

\end{document}